\documentclass[11pt]{article}
\usepackage[margin=1in]{geometry}
\usepackage[T1]{fontenc}
\usepackage[english]{babel}
\usepackage[utf8x]{inputenc}
\usepackage{ucs}
\RequirePackage[colorlinks,linkcolor=black,citecolor=blue,urlcolor=blue]{hyperref}
\usepackage{amsmath}
\usepackage{amsfonts}
\usepackage{amssymb}
\usepackage{array}
\usepackage{graphicx}
\usepackage{subfig}
\usepackage{natbib}
\usepackage{authblk}
\usepackage[]{algorithm}
\usepackage{algpseudocode}

\newcommand{\bC}{\mathbf{C}}
\newcommand{\bY}{\mathbf{Y}}

\newcommand{\bX}{\mathbf{X}}

\newcommand{\bV}{\mathbf{V}}
\newcommand{\bM}{\mathbf{M}}
\newcommand{\bS}{\mathbf{S}}
\newcommand{\bs}{\mathbf{s}}
\newcommand{\by}{\mathbf{y}}
\newcommand{\bx}{\mathbf{x}}
\newcommand{\bz}{\mathbf{z}}
\newcommand{\btheta}{\boldsymbol{\theta}}
\newcommand{\bmu}{\boldsymbol{\mu}}
\newcommand{\bSigma}{\boldsymbol{\Sigma}}

\author{Umberto Picchini \bigskip
  \\
  {\small Centre for Mathematical Sciences, Lund University},\\
  {\small S\"{o}lvegatan 18},
  {\small SE-22100 Lund, Sweden}\\
  {\small Email:} {\small {\tt umberto@maths.lth.se}}
  }

\title{Likelihood-free stochastic approximation EM for inference in complex models}

\date{}

\begin{document}

\maketitle


\begin{abstract}
A maximum likelihood methodology for the parameters of models with an intractable likelihood is introduced. We produce a likelihood-free version of the stochastic approximation expectation-maximization (SAEM) algorithm to maximize the likelihood function of model parameters. While SAEM is best suited for models having a tractable ``complete likelihood'' function, its application to moderately complex models is a difficult or even impossible task. We show how to construct a likelihood-free version of SAEM by using the ``synthetic likelihood'' paradigm. Our method is completely plug-and-play, requires almost no tuning and can be applied to both static and dynamic models. Four simulation studies illustrate the method, including a stochastic differential equation model, a stochastic Lotka-Volterra model and data from $g$-and-$k$ distributions. MATLAB code is available as supplementary material.
\end{abstract}

\noindent%
{\it Keywords:}  incomplete data; intractable likelihood; Lotka-Volterra; SAEM; stochastic differential equation; synthetic likelihood; state space model.

\section{Introduction}

Most mathematical/statistical models for realistic experiments include unobservable (latent) components $\bX$ that complicate the statistical inference for model parameters $\btheta$.
Here we consider the problem of estimating $\btheta$, given an observable process $\bY$ from which data are generated, in models characterized by missing (incomplete) data in the sense discussed in \cite{Dempster1977} when introducing the celebrated EM algorithm. 
Therefore, our goal is to estimate $\btheta$, in presence of a latent (unobservable) $\bX$ on which observed data depend. 

While here we deal with a modification of an EM-type algorithm, for the moment our interest is to discuss the inference problem for models having so-called ``intractable likelihoods''. For these models the likelihood function is unavailable in closed form and obtaining an approximation (or evaluating said approximation) is computationally prohibitive. Two of the discussed examples are state-space models (SSM), and for SSM recent advancements in sequential Monte Carlo methods (also known as particle filters) have revolutionised the practical application of statistical inference, especially the Bayesian kind, see the review in \cite{kantas2015particle}. For more general models than SSM, approximate Bayesian computation (ABC) is often the only available solution to perform statistical inference for the parameters of complex models with intractable likelihoods. ABC (see \citealp{marin-et-al(2011)} for a review) is an ensemble of algorithms that only requires the ability to generate synthetic observations from the assumed data generating model, hence these are ``plug-and-play'' algorithms. While ABC algorithms have been developed since the '90s, the most important issues for a successful implementation of ABC are still as relevant today as they were twenty years ago. In particular, the most typical usage of ABC requires the analyst to specify summary statistics that are ``informative'' regarding the unknown $\btheta$. Moreover, a threshold parameter is introduced to compare summary statistics computed on the available data with summaries computed on simulations from the assumed data generating model. The problem of selecting appropriate summaries is the most serious of the two (see \citealp{fearnhead-prangle(2011)}). The determination of the threshold for summaries comparison is also very important and has a significant impact on the computational budget. Finally, when ABC is implemented within an MCMC sampler, there is a further layer of practical issues that are usually of difficult management for the non-expert practitioner, such as coding an appropriate adaptive MCMC method for the generation of parameter proposals, also noting that the frequency of the adaptation affects results.
It is fair to say that calibration of ABC algorithms is often not trivial. 
A more recent plug-and-play methoology is given by synthetic likelihoods (SL) \citep{wood2010statistical}. SL requires the specification of data summaries, but no threshold parameter is introduced and the weighting of the summaries is automatically handled, thus the method is very easy to implement. However, while ABC sets no assumptions on said summaries, SL assumes a multivariate Gaussian distribution: hence, SL is less general than ABC and as discussed in \cite{price2016bayesian} and in section \ref{sec:lv}, significant departures from the assumed Gaussianity can have a negative impact on inference results.

In this work we consider the idea underlying the synthetic likelihood approach, and embed this into the stochastic approximation (SAEM) algorithm of \cite{Delyon1999}, for maximum likelihood inference. The resulting SAEM-SL algorithm is a likelihood-free version of SAEM which is easy to code, requires minimal tuning and appeals a general class of incomplete-data models, either ``static'' (time-independent) and dynamic models. 
Since two of our simulation studies use state-space models, our notation introduces quantities that are time-indexed, however we emphasize that the methodology is suited for dynamic models that are not SSM and also for static models, see the example in the Supplementary Material.

State-space models (SSM, \citealp{cappe2005inference}), are used in many fields, such as biology, chemistry, ecology, signal processing etc. 
We now introduce some notation. Consider a stochastic process $\{\bY_t\}_{t> t_0}$, $\bY_t\in\mathsf{Y}\subseteq \mathbb{R}^{d_y}$, which is observed at discrete sampling times $t\in\{t_1,...,t_n\}$ with $t_1>t_0\geq 0$, and we denote with $\bY_{1:n}=(\bY_1,...,\bY_n)$ the corresponding observations (data) from $\{\bY_t\}_{t> t_0}$ collected at said time points, where $\bY_{t_j}\equiv \bY_j$ for $j=1,...,n$.
Consider also a latent (unobservable) continuous-time stochastic process $\{\bX_t\}_{t\geq t_0}$, $\bX_t\in\mathsf{X}\subseteq \mathbb{R}^{d_x}$.
Process $\bX_t\sim p(\bx_t|\bx_s,\btheta_x)$ is assumed Markovian with transition densities $p(\bx_t|\bx_s,\cdot)$, $t_0\leq s<t$. 
Denote with $\bX_{1:n}=(\bX_1,...,\bX_n)$ the unobserved values for $\{\bX_t\}_{t\geq t_0}$ at times $\{t_1,...,t_n\}$ and set $\bX_{0:n}=(\bX_0,\bX_{1:n})$, where $\bX_0$ is the (random or fixed) initial state for $\{\bX_t\}$ at time $t_0$.
Both processes $\{\bX_t\}$ and $\{\bY_t\}$ depend on their own (assumed unknown) vector-parameters $\btheta_x$ and $\btheta_y$ respectively.
We think at $\{\bY_t\}$ as a measurement-error-corrupted version of $\{\bX_t\}$ and assume that observations for $\{\bY_t\}$ are conditionally independent given $\{\bX_t\}$.
The SSM can be summarised as
\begin{equation}
\begin{cases}
\bY_j\sim p(\by_j|\bX_j,\btheta_y),\quad j=1,...,n\\
\bX_t\sim p(\bx_t|\bx_s,\btheta_x),\quad t_0\leq s<t.
\end{cases}
\label{eq:state-space-general}
\end{equation}
Typically $p(\by_j|\bX_j,\cdot)$ is a known density (or probability mass) function. Regarding the transition density $p(\bx_t|\bx_s,\cdot)$, this is typically unknown except for very simple toy models.

Goal of our work is to estimate the parameters $(\btheta_x,\btheta_y)$ by maximum likelihood using data $\bY_{1:n}=(\bY_1,...,\bY_n)$. For ease of notation we refer to the vector $\btheta:=(\btheta_x,\btheta_y)$ as the object of our inference. As previously remarked, the SAEM-SL methodology we introduce does not require data generated from a SSM, hence conditional independence of observations and Markovianity of $\{\bX_t\}$ are not necessary for SAEM-SL to work.

The well-known EM algorithm \citep{Dempster1977} is suitable for maximum likelihood estimation for incomplete-data models. EM computes the conditional expectation of the complete-likelihood for the pair $(\bY_{1:n},\bX_{0:n})$ and then produces a (local) maximizer for the data likelihood function based on observations $\bY_{1:n}$. One of the difficulties with EM is to compute the conditional expectation of the state $\{\bX_t\}$ given the observations  $\bY_{1:n}$. This conditional expectation can be computed exactly with the Kalman filter when the state-space is linear and Gaussian \citep{cappe2005inference}, and otherwise it has to be approximated. In this work we focus on a stochastic approximation of the EM algorithm, namely the Stochastic Approximation EM (SAEM) \citep{Delyon1999}. The problem with implementing SAEM is at least two-fold: (i) it is necessary to generate  an appropriate ``proposal'' for the state $\{\bX_t\}$, conditionally on the current value of $\btheta$. Sequential Monte Carlo (SMC) algorithms \citep{doucet2001sequential} can provide such state proposal, and have already been coupled to stochastic EM algorithms (see e.g. \cite{HuysPaninski2009, Lindsten2013, Ditlevsen2014} and references therein). However (ii) a second and perhaps more serious difficulty is that in order to use SAEM the complete likelihood of $\btheta$ based on the joint distribution of $(\bY_{1:n},\bX_{0:n})$ must be tractable. With ``tractable'' we mean that the model at hand has a complete likelihood that it is possible to write in closed-form, and that additionally it is possible to derive analytically essential quantities, such as the corresponding sufficient statistics: this is because the convergence of SAEM to the maximizer of the data likelihood is ensured only for observations belonging to the exponential family. These requirements are usually  very difficult to satisfy, or result impossible for most realistic models. Even when these can be satisfied, the required analytic work is at best a tedious, difficult and error-prone task. Also, such difficulties force the modeller to formulate oversimplified, tractable models so that SAEM can be implemented. However realistic models call for more complex formulations which are usually not amenable to closed form analytic computations.

\section{The complete likelihood and stochastic approximation EM}
\label{sec:complete-likelihood}

Recall that $\bY_{1:n}=(\bY_1,...,\bY_n)$ denotes the available data collected at times $(t_1,...,t_n)$ and denote with $\bX_{1:n}=(\bX_1,...,\bX_n)$ the corresponding unobserved states. We additionally set $\bX_{0:n}=(\bX_0,\bX_{1:n})$ for the vector including an initial (fixed or random) state $\bX_0$, that is $\bX_1$ is generated as $\bX_1\sim p(\bx_1|\bx_0)$. When the transition densities between sampling times $p(\bx_j|\bx_{j-1})$ are available in closed form ($j=1,...,n$), the ``data likelihood'' function for $\btheta$ (sometimes denoted ``incomplete data likelihood'') can be written as 
\begin{align}
p(\bY_{1:n};\btheta) &= \int p_{\bY,\bX}(\bY_{1:n},\bX_{0:n}\, ; \btheta)\,d\bX_{0:n} = \int p_{\bY|\bX}(\bY_{1:n}|\bX_{0:n}\, ; \btheta)p_{\bX}(\bX_{0:n}; \btheta)\,d\bX_{0:n}\nonumber\\
&=\int p(\bX_0)\biggl\{\prod_{j=1}^n p(\bY_j|\bX_j;\btheta)p(\bX_j|\bX_{j-1};\btheta)\biggr\}d\bX_0\cdots d\bX_n  \label{eq:likelihood}
\end{align}
where we have assumed a random initial state with density $p(\bX_0)$.
Here $p_{\bY,\bX}$ is the ``complete data likelihood'', $p(\bY_j|\bX_j)$ the conditional density of $\bY_j$ and $p_{\bX}(\bX_{0:n}; \btheta)$ the joint density of $\bX_{0:n}$. The last equality in \eqref{eq:likelihood} exploits the notion of conditional independence of observations given latent states and the Markovian property of $\{\bX_t\}$.  
In general the likelihood \eqref{eq:likelihood} is not explicitly known either because the integral is multidimensional or because  expressions for transition densities are typically not available. 
In addition, when an exact simulator for the solution of the dynamical process associated with the Markov process $\{\bX_t\}$ is unavailable, hence it is not possible to sample from $p(\bX_j|\bX_{j-1};\btheta)$, numerical discretisation methods are required, see the example in section \ref{sec:theoph}. Without loss of generality, say that we have equispaced sampling times such that $t_j=t_{j-1}+\Delta$, with $\Delta>0$. Now introduce a discretisation for the interval $[t_1,t_n]$ given by $\{\tau_1,\tau_h,...,\tau_{Gh},...,\tau_{nGh}\}$ where $h=\Delta/G$ and $G\geq 1$. We take $\tau_1=t_1$, $\tau_{nGh}=t_n$ and therefore $\tau_{i}\in \{t_1,....,t_n\}$ for $i=1,Gh,2Gh,...,nGh$. We denote with $N$ the number of elements in the discretisation $\{\tau_1,\tau_h,...,\tau_{Gh},...,\tau_{nGh}\}$ and with $\bX_{1:N}= (\bX_{\tau_1}, \ldots, \bX_{\tau_N})$ the corresponding values of $\{\bX_t\}$ obtained when using a given numerical/approximated method of choice. Then the likelihood function becomes
\begin{align*}
p(\bY_{1:n};\btheta) &= \int p_{\bY,\bX}(\bY_{1:n},\bX_{0:N}\, ; \btheta)\,d\bX_{0:N} = \int p_{\bY|\bX}(\bY_{1:n}|\bX_{0:N}\, ; \btheta)p_{\bX}(\bX_{0:N}; \btheta)\,d\bX_{0:N}\\
&=\int \biggl\{\prod_{j=1}^n p(\bY_j|\bX_j;\btheta)\biggr\}p(\bX_0)\prod_{i=1}^N p(\bX_i|\bX_{i-1};\btheta)d\bX_0\cdots d\bX_N, 
\end{align*}
where the product having index $j$ is over the $\bX_{t_j}$'s and the product having index $i$ is over the $\bX_{\tau_i}$'s.

\subsection{The standard SAEM algorithm}\label{sec:standard-SAEM}
Let us briefly cover the EM principle \citep{Dempster1977}. 
 The complete data of the model is  $(\bY_{1:n},\bX_{0:N})$, where $\bX_{0:N}\equiv \bX_{0:n}$ if numerical discretisation is not required, and for ease of writing we denote this as $(\bY,\bX)\equiv (\bY_{1:n},\bX_{0:N})$ for the remaining of this section.
The EM algorithm maximizes the function
 $Q(\btheta|\btheta')=\mathbb{E}(L_c(\bY,\bX;\btheta)|\bY;\btheta')$ 
in two steps, where $L_c(\bY,\bX;\btheta):=\log p_{\bY,\bX}$ is the log-likelihood of the \textit{complete} data and $\mathbb{E}$ is the conditional expectation under the conditional distribution $p_{\bX|\bY} (\cdot ; \btheta')$. 
More explicitly, by denoting with $\hat{\btheta}^{(k-1)}$ the parameter estimate  obtained at iteration $k-1$ of EM, at $k$th iteration of EM the E-step  computes $Q(\btheta|\hat{\btheta}^{(k-1)})=\int \log p_{\bY,\bX}(\bY,\bX;\btheta)p_{\bX|\bY}(\bX|\bY;\hat{\btheta}^{(k-1)})d\bX$. The M-step computes $\hat{\btheta}^{(k)}=\arg\max_{\btheta\in\Theta} Q(\btheta|\hat{\btheta}^{(k-1)})$. The resulting sequence $\{\hat{\btheta}^{(k)}\}_k$ converges to a stationary point of the data likelihood $p(\bY;\btheta)$, under weak assumptions. In most cases the E-step is difficult to perform, while the M-step can be considered relatively straightforward, meaning that standard optimization procedures for the M-step can be implemented, or closed form solutions are possible. 

Important strategies for dealing with an intractable E-step are MCEM \citep{wei1990monte} and SAEM \citep{Delyon1999}, see also \cite{Lindsten2013} for a synthetic review. 
In SAEM the integral in $Q(\btheta|\hat{\btheta}^{(k-1)})$ is approximated using a stochastic procedure.
SAEM is proved to converge under general
conditions if $L_c(\bY,\bX;\btheta)$ belongs to the regular  exponential family
\begin{equation}
L_c(\bY,\bX;\btheta)= -\Lambda(\btheta)+\langle \bS_c(\bY,\bX),\Gamma(\btheta)\rangle, \label{eq:complete-likelihood}
\end{equation}
where $\left\langle ...\right\rangle$ is the scalar product, $\Lambda$ and $\Gamma$ are two functions of $\btheta$ and  $\bS_c(\bY,\bX)$ is  the minimal sufficient statistic of the complete model.  The E-step is then divided into a
simulation step (S-step) of the missing data $\bX^{(k)}$ under the
conditional distribution $p_{\bX|\bY}(\cdot;\hat{\btheta}^{(k-1)})$ and a stochastic
approximation step (SA-step) of the conditional expectation, using $(\gamma_k)_{k\geq 1}$ a sequence of real numbers in $[0,1]$, such that $\sum_{k=1}^\infty\gamma_k=\infty$ and $\sum_{k=1}^\infty\gamma_k^2<\infty$. This SA-step approximates  $\mathbb{E}\left\lbrack  \bS_c(\bY,\bX) \vert  \hat{\btheta}^{(k-1)} \right\rbrack$ at each iteration by the value $\bs_k$ defined recursively as follows
$$
\bs_{k}=\bs_{k-1}+\gamma_k( \bS_c(\bY,\bX^{(k)})-\bs_{k-1}).
$$
The M-step is thus the update of the estimates $\hat{\btheta}^{(k-1)}$
\begin{equation}
\hat{\btheta}^{(k)}= \arg \max_{\btheta \in \Theta} \left(-\Lambda(\btheta)+\langle \bs_{k},\Gamma(\btheta)\rangle \right).\label{eq:M-step}
\end{equation}
A schematic description of the SAEM procedure (coupled with a bootstrap filter) is in algorithm \ref{alg:saem-smc}, see also \cite{picchini2016coupling}. 
Moreover, when it is possible to parametrize the complete loglikelihood in terms of $\bS_c(\cdot)$ as in \eqref{eq:complete-likelihood}, then it is sometimes possible to determine the $\hat{\btheta}^{(k)}$ in \eqref{eq:M-step} explicitly (see sections \ref{sec:nonlingauss}--\ref{sec:theoph}), and this has an obvious computational advantage.

Usually, the simulation step of the hidden trajectory $\bX^{(k)}$ conditionally to the observations $\bY$ cannot be performed directly. A standard possibility is to use $M$ ``particles'' from sequential Monte Carlo filters, such as the bootstrap filter \citep{gordon1993novel}, see algorithm \ref{alg:bf}. 
\begin{algorithm}
\small
\caption{SAEM with a bootstrap filter}\label{alg:saem-smc}
\begin{algorithmic}
\State Step 0. Set parameters starting values $\hat{\btheta}^{(0)}$, then set $M$, $\bar{M}$ and $k:=1$.
\State Step 1. For fixed $\hat{\btheta}^{(k-1)}$ apply the bootstrap filter in algorithm \ref{alg:bf} with $M$ particles and particles threshold $\bar{M}$. 
\State 2 Sample an index $m'$ from the probability distribution $\{w_n^{(1)},...,w_n^{(M)}\}$ on $\{1,...,M\}$ and form the path $\bX^{(k)}$ resulting from the genealogy of $m'$.
\State Step 3. {\bf Stochastic Approximation step :} update of the sufficient statistics
	$$\bs_{k}  =  \bs_{k-1} + \gamma_k\,\left(\bS_c(\bY, \bX^{(k)}) -\bs_{k-1}\right)$$
\State Step 4. {\bf Maximisation step:} update $\btheta$
	$$\hat{\btheta}^{(k)}= \arg \max_{\btheta \in \Theta} \left(-\Lambda(\btheta)+\langle \bs_{k},\Gamma(\btheta)\rangle \right)$$
Set $k:=k+1$ and go to step 1.
\end{algorithmic}
\end{algorithm}
\begin{algorithm}
\small
\caption{Bootstrap filter}\label{alg:bf}
\begin{algorithmic}
\State Step 0. Set $j=1$: for $m=1,...,M$ sample $\bX_1^{(m)}\sim p(\bX_0)$, compute weights $W_1^{(m)}=f(\bY_1|\bX_1^{(m)})$ and normalize weights $w_1^{(m)}:=W_1^{(m)}/\sum_{m=1}^M W_1^{(m)}$.
\State Step 1.
   \If{$ESS(\{w_j^{(m)}\})<\bar{M}$} 
\State resample $M$   particles $\{\bX_j^{(m)},w_j^{(m)}\}$ and set $W_j^{(m)}=1/M$. 
   \EndIf
\State Set $j:=j+1$ and if $j=n+1$, stop and return all constructed weights $\{W_j^{(m)}\}_{j=1:n}^{m=1:M}$ to sample a single path (see main text). Otherwise go to step 2.
\State Step 2. For $m=1,...,M$ sample $\bX_j^{(m)}\sim p(\cdot|\bX_{j-1}^{(m)})$. Compute 
$$W_j^{(m)}:=w_{j-1}^{(m)}p(\bY_j|\bX_j^{(m)})$$
normalize weights $w_j^{(m)}:=W_{j}^{(m)}/\sum_{m=1}^M W_j^{(m)}$ and go to step 1.
\end{algorithmic}
\end{algorithm}
The quantity ESS in algorithm \ref{alg:bf} is the effective sample size (e.g. \citealp{liu2008monte}) often estimated as $ESS(\{w_j^{(m)}\})=1/\sum_{m=1}^M(w_j^{(m)})^2$ and taking values between 1 and $M$, while $\bar{M}\leq M$ is a  threshold value that ``activates'' the resampling step, see \cite{Cappe2007} for an introduction to particle filters.
In addition to the procedure outlined in algorithm \ref{alg:bf}, once the set of normalised weights $\{w_n^{(1)},...,w_n^{(M)}\}$ is available at the end of the bootstrap filter, we sample a single index from the set $\{1,...,M\}$ having associated probabilities $\{w_n^{(1)},...,w_n^{(M)}\}$. Denote with $m'$ such index and with $a_j^m$ the ``ancestor'' of the generic $m$th particle sampled at time $t_{j+1}$, with $1\leq a_j^m \leq M$ ($m=1,...,M$, $j=1,...,n$). Then we have that particle $m'$ has ancestor $a_{n-1}^{m'}$ and in general particle $m''$ at time $t_{j+1}$ has ancestor $b_j^{m''}:=a_j^{b^{m''}_{j+1}}$, with $b_n^{m'}:=m'$. Hence, at the end of algorithm \ref{alg:bf} we can sample $m'$ and construct its genealogy (see also \citealp{andrieu2010particle}): the sequence of states $\{\bX_t\}$ resulting from the genealogy of $m'$ is the chosen path that will be passed to SAEM in algorithm 
\ref{alg:saem-smc}.

However, as explained in the Introduction and self-evident in the application in section \ref{sec:theoph}, constructing the SAEM machinery is a challenging task for most realistic models as typically the sufficient statistics $\bS_c$ for the complete loglikelihood need to be available, for computational efficiency. Moreover, for state-space models it is necessary to know the expression of the transition densities, to construct the complete loglikelihood. For most stochastic nonlinear models, transition densities are typically unavailable in closed form. Finally, even when SAEM is implemented for state-space models, as highlighted in \cite{picchini2016coupling} the particles selected from the bootstrap filter might result in a poor estimation when the resampling step is frequently triggered (see \citealp{picchini2016coupling} for solutions).

In section \ref{sec:synlik} we propose a new, likelihood-free version of SAEM, that is not restricted to dynamic models. But first, it is necessary to introduce the synthetic likelihoods methodology, due to \cite{wood2010statistical}.

\section{Synthetic likelihoods}\label{sec:synlik}

Same as for approximate Bayesian computation (ABC) algorithms, synthetic likelihoods \citep{wood2010statistical} is an ``information reduction strategy'' that constructs inference based on a set of ad-hoc summaries of the data $\bS(\bY)$, rather than use the full dataset $\bY$ directly. These summaries are defined by the analyst and have nothing to do with the complete sufficient summaries $\bS_c$ in \eqref{eq:complete-likelihood}. The synthetic likelihoods methodology assumes the data summaries to be jointly multivariate Gaussian as $\bS(\bY)\sim \mathcal{N}(\bmu_{\btheta},\bSigma_{\btheta})$, with unknown mean $\bmu_{\btheta}$ and unknown covariance matrix $\bSigma_{\btheta}$. Instead, ABC does not make any parametric assumption on the summaries. Notation-wise we make explicit the dependence of the mean and covariance on $\btheta$, as later on it will be important to highlight this fact when estimating $\btheta$ (e.g. in equation \eqref{eq:synlik-maximization}). 

Estimators for $\bmu_{\btheta}$ and $\bSigma_{\btheta}$ are found by simulating $R$ datasets independently from the assumed data-generating model, conditionally on some $\btheta$. 
We denote the artificial datasets simulated from model \eqref{eq:state-space-general} with $\by_1^{*},...,\by_R^{*}$. These are such that $\dim(\by^*_r)=\dim(\bY)$, $r=1,...,R$. For each dataset \cite{wood2010statistical} constructs the corresponding (vector valued) summary $\bS_r^*$, with $\dim(\bS_r^{*})=\dim(\bS(\bY))$. Then he computes the following estimators:
\[
\hat{\bmu}_{\btheta} = \frac{1}{R}\sum_{r=1}^{R}\bS_r^{*},\qquad  
\hat{\bSigma}_{\btheta} = \frac{1}{R-1}\sum_{r=1}^{R}(\bS_r^{*}-\hat{\bmu}_{\btheta})(\bS_r^{*}-\hat{\bmu}_{\btheta})'.
\]
A ``synthetic likelihood'' based on the summaries for the observed data is defined as $\hat{p}(\bS(\bY)|\btheta):=\mathcal{N}(\bS(\bY);\hat{\bmu}_{\btheta},\hat{\bSigma}_{\btheta})$. It is then possible to numerically maximize $\hat{p}(\bS(\bY)|\btheta)$ with respect to $\btheta$ or compute the MAP (maximum a posteriory) for the associated posterior distribution using MCMC, by using uniform priors for the parameters. In order to construct synthetic likelihoods the only parameter that needs to be set is $R$ (we consider the statistics $\bS(\cdot)$ as part of the model specification).

\section{SAEM with synthetic likelihoods }\label{sec:saem-sl}

We now use synthetic likelihoods (SL) to develop a likelihood-free version of SAEM. The main consequences of our approach are (i) sufficient statistics for the complete (synthetic) likelihood are immediately available, via simulation; (ii) we allow the SAEM optimizer to be implemented for complex/intractable models and (iii) the algorithm does not require advanced tuning. With specific reference to existing synthetic likelihoods approaches, with SAEM-SL the user does not need to set-up an MCMC implementation, as instead required in \cite{wood2010statistical} and \cite{price2016bayesian} and this usually comes with a need for expert tuning, as discussed in the introduction. A disadvantage of SAEM-SL is that uncertainty quantification is not provided. Denote with $\bS(\bY)$ and $\bS(\bX)$ \textit{user-defined} summary statistics for $\bY$ and $\bX$ respectively. Again, these are meant to encode information regarding $\btheta$.
Define $\bS=(\bS(\bY),\bS(\bX))$ and assume the complete likelihood for $\bS$ to be a multivariate Gaussian with mean $\bmu_{\btheta}$ and covariance $\bSigma_{\btheta}$. That is for the corresponding ``complete synthetic log-likelihood'' evaluated at $\bS$ we set
\begin{equation}
L_c(\bS;\btheta):=L_c(\bS(\bY),\bS(\bX);\btheta)=\log \mathcal{N}(\bS;\bmu_{\btheta},\bSigma_{\btheta}).
\label{eq:gauss-complete}
\end{equation}
Of course $\bmu_{\btheta}$ and $\bSigma_{\btheta}$ are in general unknown. Also, here $\bmu_{\btheta}$ and $\bSigma_{\btheta}$ are not the same moments defined for the data likelihood in section \ref{sec:synlik}, as the latter is based solely on $\bS(\bY)$.

Here we illustrate an instance of SL for the current $\btheta$, this returning estimators $\hat{\bmu}_{\btheta}$ and $\hat{\bSigma}_{\btheta}$. We call this procedure ``internal SAEM-SL'' to be distinguished from an ``external'' procedure described later. Crucially,
 thanks to the Gaussian assumption set on the user's summaries $\bS$ it is known that $(\hat{\bmu}_{\btheta},\hat{\bSigma}_{\btheta})$ is jointly sufficient for $({\bmu}_{\btheta},{\bSigma}_{\btheta})$. Hence we are allowed to set the following equality for the complete sufficient statistics ${\bS}_c(\bS(\bY),\bS(\bX)):=(\hat{\bmu}_{\btheta},\hat{\bSigma}_{\btheta})$ without the need to perform analytic calculations. Then we plug the obtained moment estimates into the ``external SAEM-SL''. While below we describe the several steps of our approach, the complete procedure is illustrated in algorithm \ref{alg:saem-sl}. 

\subsection*{Internal SAEM-SL}
Assume a value for $\btheta$ is given.
\begin{enumerate}
\item Simulate independently from the model $R$ realizations of processes $\{\bX_t\}$ and $\{\bY_t\}$: $\bx_r^*\sim p_\bX(\bX_{0:N};\btheta)$ and $\by_r^*\sim p_{\bY|\bX}(\bY_{1:n}|\bx_r^*;\btheta)$, $r=1,...,R$.
\item compute user-defined summaries $\bS_r^*=(\bS(\by_r^*),\bS(\bx_r^*))$ for each $r$.
\item estimate moments (sufficient statistics for $(\bmu_{\btheta},\bSigma_{\btheta})$)
\begin{equation}
\hat{\bmu}_{\btheta} = \frac{1}{R}\sum_{r=1}^{R}\bS_r^{*},\qquad  
\hat{\bSigma}_{\btheta} = \frac{1}{R-1}\sum_{r=1}^{R}(\bS_r^{*}-\hat{\bmu}_{\btheta})(\bS_r^{*}-\hat{\bmu}_{\btheta})'\label{eq:synthetic-sample-moments}.
\end{equation}
\end{enumerate}

\subsection*{External SAEM-SL}

A generic iteration of SAEM is executed using the estimators $(\hat{\bmu}_{\btheta},\hat{\bSigma}_{\btheta})$ from \eqref{eq:synthetic-sample-moments}. At iteration $k$ we update separately the moments for the complete loglikelihood as 
\begin{align}
\hat{\bmu}^{(k)}_{\btheta}&=\hat{\bmu}^{(k-1)}_{\btheta}+\gamma^{(k)}( \hat{\bmu}_{\btheta}-\hat{\bmu}^{(k-1)}_{\btheta})\label{eq:update-mean}\\
\hat{\bSigma}^{(k)}_{\btheta}&=\hat{\bSigma}^{(k-1)}_{\btheta}+\gamma^{(k)}( \hat{\bSigma}_{\btheta}-\hat{\bSigma}^{(k-1)}_{\btheta})\label{eq:update-cov}.
\end{align}
From the quantities computed in \eqref{eq:update-mean}-\eqref{eq:update-cov} extract the corresponding mean and covariances for the two simulated processes, that is set $\hat{\bmu}^{(k)}\equiv \hat{\bmu}^{(k)}_{\btheta}=(\hat{\bmu}_{x},\hat{\bmu}_{y})$ and 
\[
\hat{\bSigma}^{(k)}\equiv \hat{\bSigma}^{(k)}_{\btheta} = \left[\begin{array}{cc}
\hat{\bSigma}_{x} & \hat{\bSigma}_{xy}   \\
\hat{\bSigma}_{yx} & \hat{\bSigma}_{y} \\
\end{array}\right].
\]
We now sample $\bS(\bX^{(k)})$ conditionally on $\bS(\bY)$ by using well known properties of Gaussian distributions: we have $\bS(\bX^{(k)})|\bS(\bY)\sim \mathcal{N}(\hat{\bmu}_{x|y,{\btheta}}^{(k)},\hat{\bSigma}_{x|y,{\btheta}}^{(k)})$ where (here we drop the index $k$ and subscript ${\btheta}$ for ease of reading) 
\begin{align}
\hat{\bmu}_{x|y} & = \hat{\bmu}_{x}+\hat{\bSigma}_{xy}\hat{\bSigma}_{y}^{-1}(\bS(\bY)-\hat{\bmu}_{y})\label{eq:condmean}\\
\hat{\bSigma}_{x|y} & = \hat{\bSigma}_{x}-\hat{\bSigma}_{xy}\hat{\bSigma}_{y}^{-1}\hat{\bSigma}_{yx}.\label{eq:condcov}
\end{align}
Some care should be used with the covariance matrix $\hat{\bSigma}_{x|y}^{(k)}$ when sampling $\bS(\bX^{(k)})|\bS(\bY)$, as such covariance must be positive semi-definite. In fact $\hat{\bSigma}_{x|y}^{(k)}$ is extracted from $\hat{\bSigma}^{(k)}$, however while it is known that a linear combination (via \eqref{eq:update-cov}) of semi-positive definite matrices is a semi-positive definite matrix and while the sample covariance created in the Internal SAEM-SL is by definition semi-positive definite, in numerical calculations it can still happen that the resulting matrix has negative eigenvalues due to round-off errors in floating point approximations. Therefore, before using $\hat{\bSigma}_{x|y}^{(k)}$ in our conditional sampling, we first check whether this is a positive definite matrix. If it turns out to be positive definite, by using the Cholesky decomposition of $\hat{\bSigma}_{x|y}^{(k)}$, then we proceed with the sampling, that is we obtain the lower triangular matrix $\bM$ such that $\bM\bM'=\hat{\bSigma}_{x|y}^{(k)}$ and then sample $\bS(\bX^{(k)})|\bS(\bY)$ using $\bS(\bX^{(k)}):=\hat{\bmu}_{x|y}^{(k)}+\bM\bz$, where $\bz$ is a vector of independent draws from the standard normal distribution. For those rare instances where it is not positive definite (and not even semi-positive definite) it is possible to compute a ``nearest
semi-positive definite matrix'' (e.g. \citealp{higham1988computing}) and use this one for the sampling. 

With the $\bS(\bX^{(k)})$ that has been sampled,  set $\bS^{(k)}:=(\bS(\bY),\bS(\bX^{(k)}))$
and compute the M-step
\begin{equation}
\hat{\btheta}^{(k)}=\arg \max_{\btheta \in \Theta} \hat{L}_c(\bS^{(k)};\btheta)=\arg \max_{\btheta \in \Theta} \log \mathcal{N}(\bS^{(k)};\bmu_{\btheta},\bSigma_{\btheta})\label{eq:synlik-maximization}
\end{equation}
where maximization is obtained numerically, for example using $L$ iterations of a Nelder--Mead simplex. Each iteration of the maximizer used for \eqref{eq:synlik-maximization} tests a different value of $\btheta$ by invoking the Internal-SL procedure, hence each call evaluates the complete synthetic loglikelihood using a different set of simulated moments $(\hat{\bmu}_{\btheta},\hat{\bSigma}_{\btheta})$ produced using the synthetic likelihoods approach.
At the end of the M-step \eqref{eq:synlik-maximization}, besides $\hat{\btheta}^{(k)}$ we also retrieve the corresponding ``optimal moments'' $(\hat{\bmu}_{\btheta},\hat{\bSigma}_{\btheta})$. Optimal moments are passed to \eqref{eq:update-mean}-\eqref{eq:update-cov} for a further iteration of the External SAEM-SL procedure. Algorithm \ref{alg:saem-sl} details a single iteration of the SAEM-SL procedure, which should be executed for $k=1,...,K$ iterations, with quantities having $k=0$ denoting input/starting values. The generality of the algorithm implies that to implement all our case studies we did not need to produce significant changes to our test code.

We initialize algorithm \ref{alg:saem-sl} by setting $\bmu^{(0)}$ and $\bSigma^{(0)}$ to a vector of zeros and to a diagonal matrix with positive entries $\delta \mathbf{I}_d$ respectively, with $\delta=10^{-12}$ and $\mathbf{I}_d$ the $d$-dimensional identity matrix with $d$ the length of vector $(\bS(\bX),\bS(\bY))$. Notice that each time a numeric maximizer evaluates \eqref{eq:synlik-maximization} for the current candidate parameters $\btheta^c$ the vector $\bS^{(k)}$ does not vary within the Internal SL: $\bS^{(k)}$ contains both the observed summaries and the summaries for the latent state $\bS(\bX^{(k)})$, which should not be altered when \eqref{eq:synlik-maximization} is executed. Also notice that while in step 2 of the Internal-SL procedure the quantity $\bS(\bx_r^*)$ is computed from the user defined set of summaries, the $\bS(\bX^{(k)})$ that is plugged into $\bS^{(k)}$ is instead sampled from a multivariate Gaussian distribution.

For the sake of discussion, here we illustrate an ideal scenario which in practice cannot be attained for most realistic models, namely assuming that (a) the user defined sumaries $\bS=(\bS(\bY),\bS(\bX))$ are jointly sufficient statistics for $\btheta$, and that (b) $\bS$ is distributed according to a multivariate Gaussian, though (b) is much easier to obtain than (a). Then under (a)--(b) SAEM-SL does not result in any approximation and converges to a (local) maximizer of the data likelihood function under the same assumptions set for SAEM in \cite{Delyon1999}. In fact, if $\bS$ is sufficient for $\btheta$ then it encodes the same amount of information regarding $\btheta$ as the couple $(\bY,\bX)$, hence $\bS_c(\bY,\bX)\equiv \bS_c(\bS(\bY),\bS(\bX))$. Then, under the additional Gaussian assumption, we have $\bS_c(\bY,\bX)\equiv \bS_c(\bS(\bY),\bS(\bX))=(\hat{\bmu}_{\btheta},\hat{\bSigma}_{\btheta})$. Therefore, since the synthetic complete loglikelihood \eqref{eq:gauss-complete} is  a member of the exponential family and can thus be written as \eqref{eq:complete-likelihood}, the two assumptions for the ``ideal'' scenario fit within the SAEM approach in section \ref{sec:standard-SAEM}. Even if the two assumptions (a)--(b) are met, deviations from what is expected from the theory is due to the non-availability of an explicit M-step, as with SAEM-SL \eqref{eq:synlik-maximization} has to be solved numerically. Hence, for a finite computational budget we might not really obtain the exact maximizer from the M-step.

\begin{algorithm}
\small
\caption{A single iteration of SAEM-SL}\label{alg:saem-sl}
\begin{algorithmic}
\State \textbf{Input:} observed summaries $\bS(\bY)$, positive integers $L$ and $R$. Values for $\hat{\btheta}^{(k-1)}$, $\hat{\bmu}^{(k-1)}$ and $\hat{\bSigma}^{(k-1)}$.
\State \textbf{Output:} $\hat{\btheta}^{(k)}$.\\ 

\State At iteration $k$ of External SAEM-SL:
\State 1. Extract $\hat{\bmu}_{x}$, $\hat{\bmu}_{y}$, $\hat{\bSigma}_{x}$, $\hat{\bSigma}_{y}$, $\hat{\bSigma}_{xy}$ and $\hat{\bSigma}_{yx}$ from $\hat{\bmu}^{(k-1)}$ and $\hat{\bSigma}^{(k-1)}$. Compute conditional moments  $\hat{\bmu}_{x|y}$, $\hat{\bSigma}_{x|y}$ using \eqref{eq:condmean}--\eqref{eq:condcov}.
\State 2. Sample $\bS(\bX^{(k-1)})|\bS(\bY)\sim \mathcal{N}(\hat{\bmu}_{x|y}^{(k-1)},\hat{\bSigma}_{x|y}^{(k-1)})$ and form $\bS^{(k-1)}:=(\bS(\bY),\bS(\bX^{(k-1)}))$. 
\State 3. Obtain $(\btheta^{(k)},\bmu^{(k)},\bSigma^{(k)})$ from \texttt{InternalSL($\bS^{(k-1)},\hat{\btheta}^{(k-1)},R$)} starting at $\hat{\btheta}^{(k-1)}$.
\State 4. Increase $k:=k+1$ and go to step 1.
\\\hrulefill\\

\textbf{Function \texttt{InternalSL($\bS^{(k-1)},\btheta^{(k-1)},R$)}:}\\
\State \textbf{Input:} $\bS^{(k-1)}$, starting parameters $\btheta^{(k-1)}$, a positive integer $R$. Functions to compute simulated summaries $\bS(\by^*)$ and $\bS(\bx^*)$ must be available.  
\State \textbf{Output:} the best found $\btheta^{*}$ maximizing $\log \mathcal{N}(\bS^{(k)};\hat{\bmu},\hat{\bSigma})$ and corresponding $(\bmu^*,\bSigma^*)$.\\
  
\State Here $\btheta^c$ denotes a generic candidate value. Initially is is set to $\btheta^c:=\btheta^{(k-1)}$. 
\State i. Simulate $\bx_r^*\sim p_\bX(\bX_{0:N};\btheta^c)$, $\by_r^*\sim p_{\bY|\bX}(\bY_{1:n}|\bx^*_r;\btheta^c)$ for $r=1,...,R$.
\State ii. Compute user-defined summaries $\bS_r^*=(\bS(\by_r^*),\bS(\bx_r^*))$ for $r=1,...,R$. Construct the corresponding $(\hat{\bmu},\hat{\bSigma})$.
\State iii. Evaluate $\log \mathcal{N}(\bS^{(k)};\hat{\bmu},\hat{\bSigma})$. \\
Use a numerical procedure that performs (i)--(iii) $L$ times for different candidates $\btheta^c$ to find the best $\btheta^*$ maximizing $\log \mathcal{N}(\bS^{(k)};\hat{\bmu},\hat{\bSigma})$.
Denote with $({\bmu}^*,\hat{\bSigma}^*)$ the simulated moments corresponding to the best found $\btheta^*$. Set  $\btheta^{(k)}:=\btheta^*$. 
\State iv. Update moments: 
\begin{align*}
\hat{\bmu}^{(k)}&=\hat{\bmu}^{(k-1)}+\gamma^{(k)}( \hat{\bmu}^*-\hat{\bmu}^{(k-1)})\\
\hat{\bSigma}^{(k)}&=\hat{\bSigma}^{(k-1)}+\gamma^{(k)}( \hat{\bSigma}^*-\hat{\bSigma}^{(k-1)}).
\end{align*}
Return $(\btheta^{(k)},\hat{\bmu}^{(k)},\hat{\bSigma}^{(k)})$.
\end{algorithmic}
\end{algorithm}

The advantages of the proposed method, which we call SAEM-SL (SAEM using synthetic likelihoods) are that (i) unlike the ``standard'' SAEM, SAEM-SL is completely plug-and-play, only the ability to simulate from the model is required; (ii) while SAEM has been (perhaps exclusively?) applied to dynamic models since SMC methods are available to simulate $\bX^{(k)}|\bY$, SAEM-SL is easily applicable also to static models.  The disadvantage with SAEM-SL is the requirement to specify a set of summaries $\bS=(\bS(\bY),\bS(\bX))$ and that for each iteration of SAEM-SL the maximization of the loglikelihood \eqref{eq:synlik-maximization} consists of an iterative procedure. On the other hand SAEM-SL  considerably expands the set of problems that is possible to treat with SAEM. The standard SAEM itself is unable to deal with complex models, unless it is possible to derive the necessary constructs (sufficient statistics for the complete likelihood and corresponding updating equations for the M-step), which is often a difficult and tedious task. If the model has an intractable complete likelihood, the task is actually impossible.

\section{Simulation studies}\label{sec:simulations}
Simulations were coded in MATLAB (except for examples using the R \texttt{pomp} package) and executed on a Intel Core i7-4790 CPU 3.60 GhZ. In SAEM we always set $\gamma_k=1$ for the first $K_1$ iterations and $\gamma_k=(k-K_1)^{-1}$ for $k\geq K_1$ as in \cite{lavielle2014mixed}. However, we found that small modifications to this setup do not affect results significantly, that is using $\gamma_k=(k-K_1)^{-\beta}$ for $k\geq K_1$ and some $\beta \in (0.5,1]$ is also valid.
The numerical maximization of \eqref{eq:synlik-maximization} is performed using the Nelder-Mead simplex as implemented in the \textsc{Matlab} function \texttt{fminsearch}.
We compare our results with state-of-art algorithms for Bayesian and ``classical'' inference. MATLAB code is available at \url{https://github.com/umbertopicchini/SAEM-SL}.

\subsection{Non-linear Gaussian state-space model}\label{sec:nonlingauss}

Here we study a simple non-linear model, useful to introduce the methods. We use a setup similar to \cite{jasra2012filtering}. See also \cite{picchini2016coupling} for inference using algorithm \ref{alg:saem-smc} as well as SAEM coupled with an ABC filter. 
We have 
\begin{align}
\begin{cases}
Y_j = X_j + \sigma_y\nu_j,\qquad j\geq 1\\
X_j = 2\sin(e^{X_{j-1}})+\sigma_x\tau_j, 
\end{cases}
\label{eq:nonlingauss-ssm}
\end{align}
with $\nu_j,\tau_j\sim N(0,1)$ i.i.d. and $X_0=0$. We assume  $\sigma_x,\sigma_y>0$ as the only unknowns and therefore conduct inference for $\btheta = (\sigma^2_x,\sigma^2_y)$.
We first consider the standard SAEM methodology outlined in section \ref{sec:standard-SAEM}, and therefore construct the set of sufficient statistics corresponding to the complete log-likelihood $L_c(\bY,\bX)$. For this model the task is simple since $Y_j|X_j\sim \mathcal{N}(X_j,\sigma^2_y)$ and $X_j|X_{j-1}\sim \mathcal{N}(2\sin(e^{X_{j-1}}),\sigma^2_x)$ and it is easy to show that $S_{\sigma^2_x}=\sum_{j=1}^n (X_j-2\sin(e^{X_{j-1}}))^2$ and $S_{\sigma^2_y}=\sum_{j=1}^n (Y_j-X_{j})^2$ are sufficient for $\sigma^2_x$ and $\sigma^2_y$ respectively. By plugging these statistics into $L_c(\bY,\bX)$ and equating to zero the gradient of $L_c$ with respect to $(\sigma^2_x,\sigma^2_y)$, we find that the M-step of SAEM results in updated values for $\sigma^2_x$ and $\sigma^2_y$ given by $S_{\sigma^2_x}/n$ and $S_{\sigma^2_y}/n$ respectively. 
In the following, we write SAEM-SMC to refer to Algorithm \ref{alg:saem-smc}. 

We generate $n=50$ observations for $\{Y_j\}$ using model \eqref{eq:nonlingauss-ssm}  with $\sigma^2_x=\sigma^2_y=5$. Our setup consists in running 30 independent experiments with SAEM-SMC: for each experiment we simulate parameter starting values for $(\log\sigma_x,\log\sigma_y)$ independently generated from a bivariate Gaussian distribution with mean the true value of the parameter, i.e. $(\log\sqrt{5},\log\sqrt{5})$, and diagonal covariance matrix having (2,2) on its diagonal. Hence the starting values are very spread.
We take $K_1=300$ as the number of warmup iterations (see beginning of section \ref{sec:simulations}) and use different numbers of particles $M$ in our simulation studies, see Table \ref{tab:nonlingauss}. We impose resampling when the effective sample size ESS gets smaller than $\bar{M}=200$, for any value of $M$. In summary, for all 30 simulations we use the same data and the same setup except that in each simulation we use different starting values for the parameters. Table \ref{tab:nonlingauss} reports the median of the 30 estimates and their $1^{st}-3^{rd}$ quartiles. Simulations for $\sigma_y$ converge to completely wrong values. We also experimented with $M=5,000$ using $\bar{M}=2,000$ but this does not solve the problem with SAEM-SMC, even if we let the algorithm start at the true parameter values. However, in \cite{picchini2016coupling} we learned that SAEM-SMC (this one using the bootstrap filter) is affected by ``particles impoverishment'' degrading the quality of the inference, and therefore it is better to set a very low $\bar{M}$: in fact, when using $\bar{M}=20$ with $M=1,000$ results improve sensibly, see Table \ref{tab:nonlingauss}, though estimation of $\sigma_y$ is still unsatisfactory. See \cite{picchini2016coupling} for further insight on the problem. 

\begin{table}
\centering
\begin{tabular}{lcccc}
\hline
 $(M,\bar{M})$ & (500,200) & (1000,200) & (2000,200) & (1000,20)  \\
\hline 
\multicolumn{4}{l}{$\sigma_x$ (true value 2.23)}\\
SAEM-SMC & 2.54 [2.53,2.54] & 2.55 [2.54,2.56] & 2.55 [2.54,2.56] & 1.99 [1.85,2.14] \\
IF2* & 1.26 [1.21,1.41] & 1.35 [1.28,1.41] & 1.33 [1.28,1.40] & --\\
\hline 
\multicolumn{4}{l}{$\sigma_y$ (true value 2.23)}\\
SAEM-SMC & 0.11 [0.10,0.13] & 0.06 [0.06,0.07] & 0.04 [0.03,0.04] & 1.23 [1.00,1.39]\\
IF2* & 1.62 [1.56,1.75] & 1.64 [1.58,1.67] & 1.63 [1.59,1.67] & --\\
\hline
\hline
 $R$ & 500 & 1000 & 2000  \\
\hline 
\multicolumn{4}{l}{$\sigma_x$ (true value 2.23)}\\
SAEM-SL & 1.96 [1.27,2.52] & 1.90 [1.13,2.39] & 2.07 [1.57,2.18] & -- \\
\hline 
\multicolumn{4}{l}{$\sigma_y$ (true value 2.23)}\\
SAEM-SL & 2.35 [1.40,2.77] & 1.94 [1.30,2.44] & 1.70 [1.44,2.22]& -- \\
\hline
\end{tabular}
\caption{\footnotesize{Non-linear Gaussian model: medians and $1^{st}-3^{rd}$ quartiles for estimates obtained on 30 independent simulations, using different number of particles $M$ and different methods. (*)The IF2 method resamples at every time point, while SAEM-SMC resamples only when $ESS<\bar{M}$. Hence for IF2 it is always $\bar{M}\equiv
M$. }} 
\label{tab:nonlingauss}
\end{table}

\begin{figure}
\centering
\includegraphics[scale=0.6]{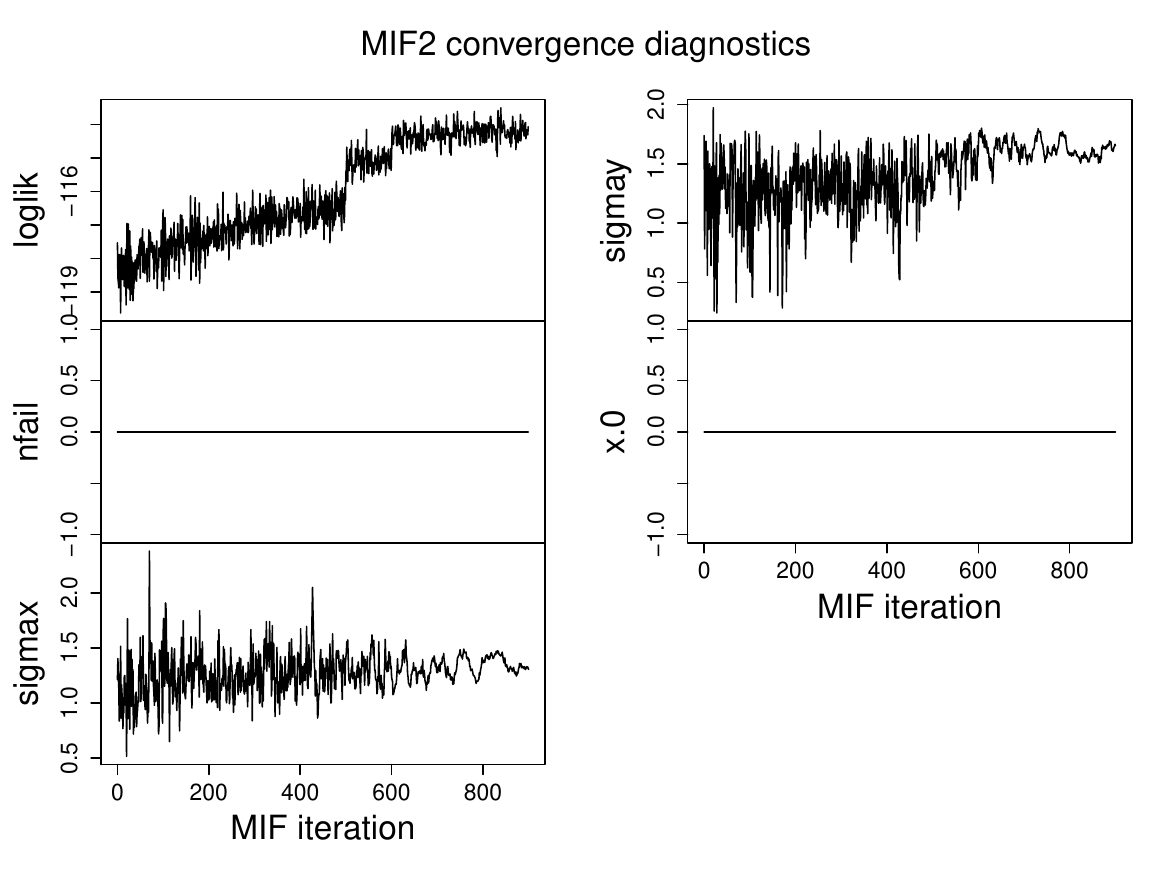}
\caption{\footnotesize{Non-linear Gaussian model: traces obtained for a single experiment with IF2 when using $M=1,000$ particles. (Top left) evolution of the loglikelihood function; (bottom left) evolution of $\sigma_x$; (top right) evolution of $\sigma_y$.}}
\label{fig:nonlingauss-IF2}
\end{figure}

We now compare the results above with the iterated filtering IF2 \citep{ionides2015inference} using the R package \texttt{pomp}. We do not provide a detailed description of IF2 here: it suffices to say that in IF2 particles are generated for both $\btheta$ (e.g. via perturbations using random walks) and for the systems state (using the bootstrap filter). Moreover a ``temperature'' parameter (to use an analogy with the simulated annealing optimization method) is let decrease until the algorithm ``freezes'' around an approximated MLE. This parameter that here we denote with $\rho$ is let decrease in $\rho\in\{0.9,0.7,0.4,0.3,0.2\}$ where the first value is used for the first 500 iterations of IF2, then   each of the remaining values is used for 100 iterations, for a total of 900 iterations. Notice that the tested version of \texttt{pomp} (v. 1.4.1.1) uses a bootstrap filter that resamples at each time point, and therefore results obtained with IF2 are not directly comparable with SAEM-SMC, hence the asterisk in Table \ref{tab:nonlingauss}. The output from one of the experiments obtained with $M=1,000$ is in Figure \ref{fig:nonlingauss-IF2}. From Figure \ref{fig:nonlingauss-IF2} we notice that the last major improvement for the loglikelihood maximization takes place  at iteration 600 when $\rho$ becomes $\rho=0.7$, and reducing $\rho$ further does not give any significant benefit (we have verified this in a number of experiments with this model), therefore we are confident about our setup. With IF2 the estimation of $\sigma_y$ is much improved compared to SAEM-SMC, however inference for $\sigma_x$ is more biased than with SAEM-SMC.

We now consider a particle marginal method (PMM, \citealp{andrieu2009pseudo}) on a single simulation (instead of thirty), as PMM is a full Bayesian methodology and results are not directly comparable with SAEM nor IF2. Once more we make use of tools provided in \texttt{pomp}. We set wide uniform priors $U(0.1,15)$ for both $\sigma_x$ and $\sigma_y$ and use $M=2,000$ particles. Also, we set the algorithm in the most favourable way, by starting it at the true parameter values (here we are only interested in using PMM to obtain exact Bayesian inference, not as a competitor to the other frequentist approaches we have illustrated). 
Parameters are proposed using an adaptive MCMC algorithm, and the algorithm is tuned to achieve the optimal 7\% acceptance rate \citep{sherlock2015efficiency}. We obtained the following posterior means and 95\% intervals: $\hat{ \sigma}_x=1.46$ [0.49,2.46], $\hat{ \sigma}_y=1.61$ [0.49,2.40]. Therefore, PMM seems to return values not very different from the ranges provided by IF2.

\begin{figure}
\centering
\includegraphics[width=17cm,height=7cm]{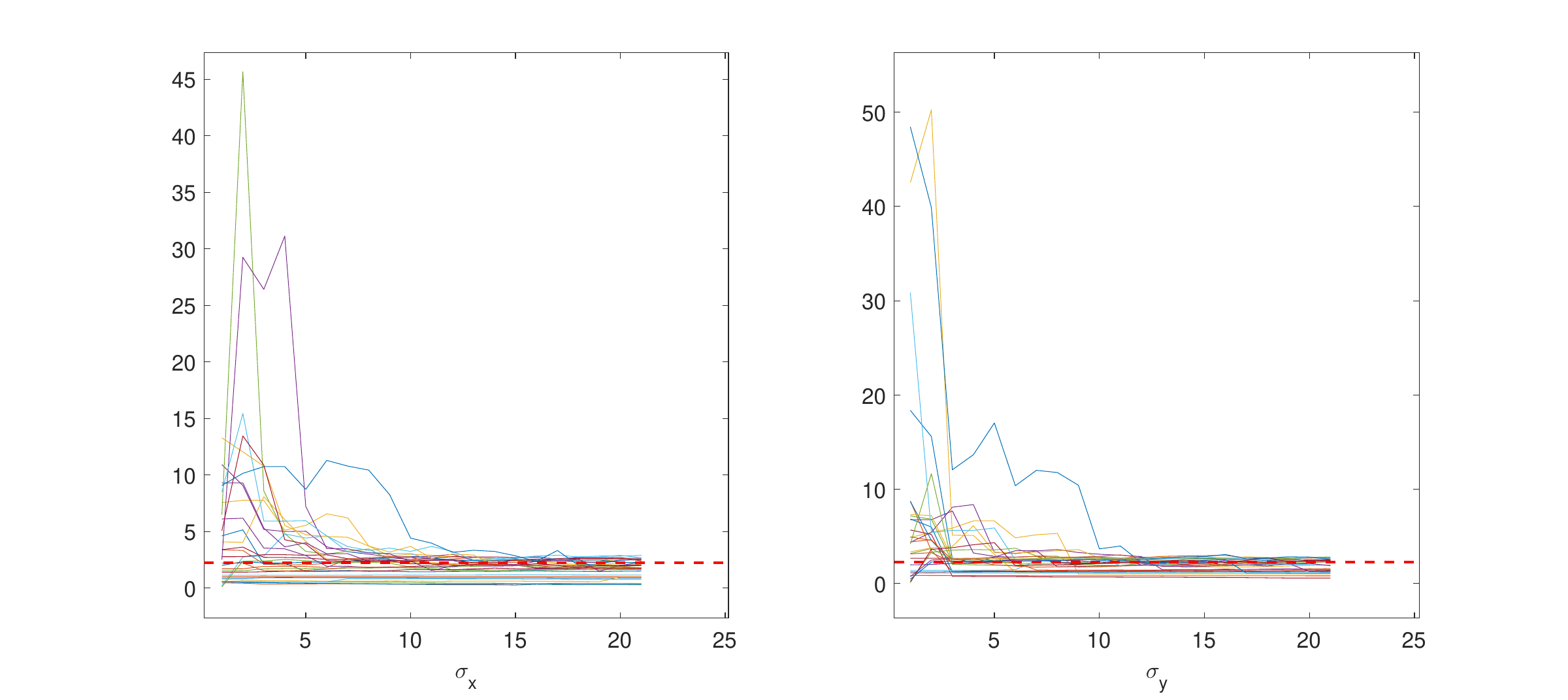}
\caption{\footnotesize{Non-linear Gaussian model: trace plots for SAEM-SL ($\sigma_x$, left; $\sigma_y$, right) when $R=1,000$ for the thirty estimation procedures. Dashed lines denote the true parameter values.}}
\label{fig:nonlingauss-traceplots_SAEM-SL}
\end{figure}

Finally, we consider inference with SAEM-SL. We performed simulations using $R=500$, 1,000 and 2,000 simulated summaries and $L=40$ iterations for the numerical maximization step. We used the same data as for SAEM-SMC and IF2, however we decide to make the estimation procedure more challenging, so we let the parameter start at random locations sampled from a Gaussian centred at $(\log \sqrt{16},\log \sqrt{16})$ and having diagonal covariance with variances $(2,2)$. Here we need to set a vector of summaries $(\bS(\bx_r^*),\bS(\by_r^*))$. Vector $\bS(\bx_r^*)$ contains (i) the median value of $\bx_r^*$; (ii) the median absolute deviation of $\bx_r^*$ and (iii)  the 10th, 20th, 75th and 90th percentile of $\bx_r^*$. Vector $\bS(\by_r^*)$ contains the same summary functions, except that these are applied to $\by_r^*$. Of course summary functions for observed data $\bS(\by):=\bS(\bY)$ are the same functions considered for $\bS(\by^*_r)$ except that now they are evaluated at $\by$.
Same as before we consider thirty repetitions of our experiment: for each experiment we run a warmup of $K_1=10$ iterations and a total number of $K=20$ SAEM-SL iterations. Results are in Table \ref{tab:nonlingauss} and trace plots for the case $R=1,000$ are in Figure \ref{fig:nonlingauss-traceplots_SAEM-SL}. As from Figure \ref{fig:nonlingauss-traceplots_SAEM-SL} we notice that those parameters initialized at much higher values than the true parameter values decay rapidly to approach the true values. As shown in Table \ref{tab:nonlingauss}, the majority of them converges to reasonable values. SAEM-SL produces excellent inference for all tested values of $R$, and convergence is very rapid, well within 10 iterations, corresponding to about 10 seconds on a computer desktop when $R=1,000$. 

\begin{figure}
\centering
\includegraphics[width=15cm,height=8cm]{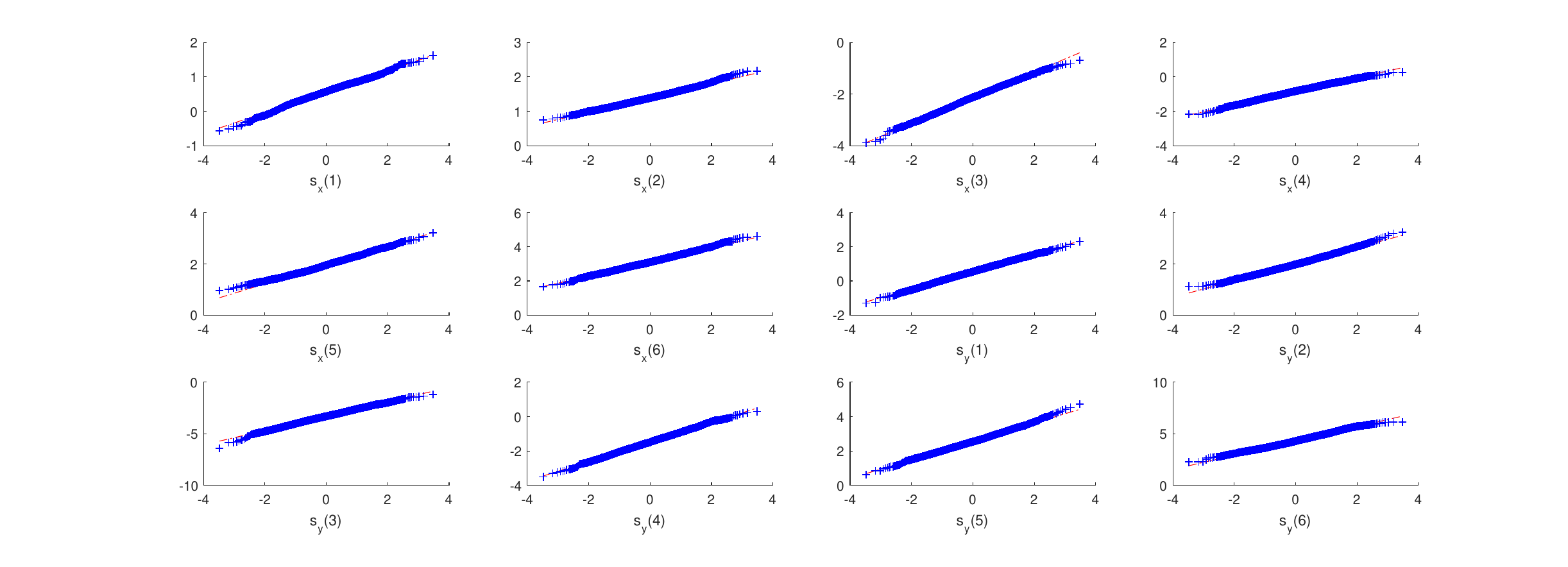}
\caption{\footnotesize{Non-linear Gaussian model: normal qq-plots for $\bS(\bx_r)$ and $\bS(\by_r)$ corresponding to the last iteration of SAEM-SL when $M=2,000$.}}
\label{fig:nonlingauss-qqplots}
\end{figure}

For  one of the thirty repetitions, Figure \ref{fig:nonlingauss-qqplots} shows the normal qq-plots for the twelve chosen summary statistics (the six statistics in $\bS(\bx_r)$ and the six in $\bS(\by_r)$) for the case $R=2,000$, generated at the optimum returned by SAEM-SL. Clearly there are no major departures from normality. Interestingly, we reach the same conclusion for the case $R=500$ (plots not reported).

\subsection{A pharmacokinetics model}\label{sec:theoph}

Here we consider a model for pharmacokinetics dynamics. For example, we may imagine to study the Theophylline drug pharmacokinetics, e.g. \cite{pinheiro1995approximations}. It will be evident that in order to apply a standard SAEM it is required some preliminary analytic effort from the modeller.
We denote with $X_t$ the level of Theophylline drug concentration in blood at time $t$ (hrs).
Consider the following non-authonomous stochastic differential equation (SDE):
\begin{equation}
dX_t = \biggl(\frac{Dose\cdot K_a \cdot K_e}{Cl}e^{-K_a t}-K_eX_t\biggr)dt + \sigma \sqrt{X_t} dW_t, \qquad t\geq t_0
\label{eq:theophylline-sde}
\end{equation}
where $Dose$ is the known drug oral dose received by a subject, $K_e$ is the elimination rate constant, $K_a$ the absorption rate constant, $Cl$ the clearance of the drug and $\sigma$ the intensity of intrinsic stochastic noise.
We simulate data measured at $n=30$ equispaced sampling times $\{t_1,t_{\Delta},...,t_{30\Delta}\}=\{1,2,...,30\}$ where $\Delta=t_j-t_{j-1}=1$. The drug oral dose is chosen to be 4 mg.
After the drug is administered, we consider as $t_0=0$ the time when the concentration first reaches $X_{t_0}=X_0=8$. The error model is assumed to be linear, $Y_j=X_j+\varepsilon_j$ where the $\varepsilon_j\sim N(0,\sigma_{\varepsilon}^2)$ are i.i.d., $j=1,...,30$. Inference is based on data $\{Y_1,...,Y_{30}\}$ collected at corresponding sampling times. Parameter $K_a$ is assumed known as it is not possible to determine the sufficient statistic for $K_a$ analytically, hence parameters of interest are $\btheta=(K_e,Cl,\sigma^2,\sigma_{\varepsilon}^2)$ as $X_0$ is also assumed known.

Equation \eqref{eq:theophylline-sde} has no available closed-form solution, hence simulated data are created in the following way. We first simulate numerically a solution to \eqref{eq:theophylline-sde} using the Euler--Maruyama discretization with stepsize $h=0.05$ on the time interval $[t_0,30]$. The Euler-Maruyama scheme is defined as
\[
X_{t+h} = X_{t} + \biggl(\frac{Dose\cdot K_a \cdot K_e}{Cl}e^{-K_a t}-K_eX_t\biggr)h + \sigma \sqrt{X_t}Z_{t+h},
\]
where the $\{Z_t\}$ are i.i.d. $\mathcal{N}(0,h)$ distributed. The grid of generated values $\bX_{0:N}$ is then linearly interpolated at sampling times $\{t_1,...,t_{30}\}$ to give $\bX_{1:n}$, and finally residual error is added to $\bX_{1:n}$ according to the error model $Y_j=X_j+\varepsilon_j$ as explained above. Data $\{Y_j\}$ are conditionally independent given the latent process $\{X_t\}$ and are generated with $(K_e,K_a,Cl,\sigma^2, \sigma_{\varepsilon}^2)=(0.05,1.492,0.04,0.01,0.102)$. The construction of the sufficient statistics to implement the standard SAEM approach is given in the Supplementary Material, and this should make evident how applying SAEM can be laborious, even for a one-dimensional model. In the results section below we show the simplicity of application of SAEM-SL for this specific example and compare SAEM-SL with a number of alternative approaches.

\subsubsection{Results}\label{sec:theophylline-results}

\begin{figure}
\centering
\includegraphics[scale=0.6]{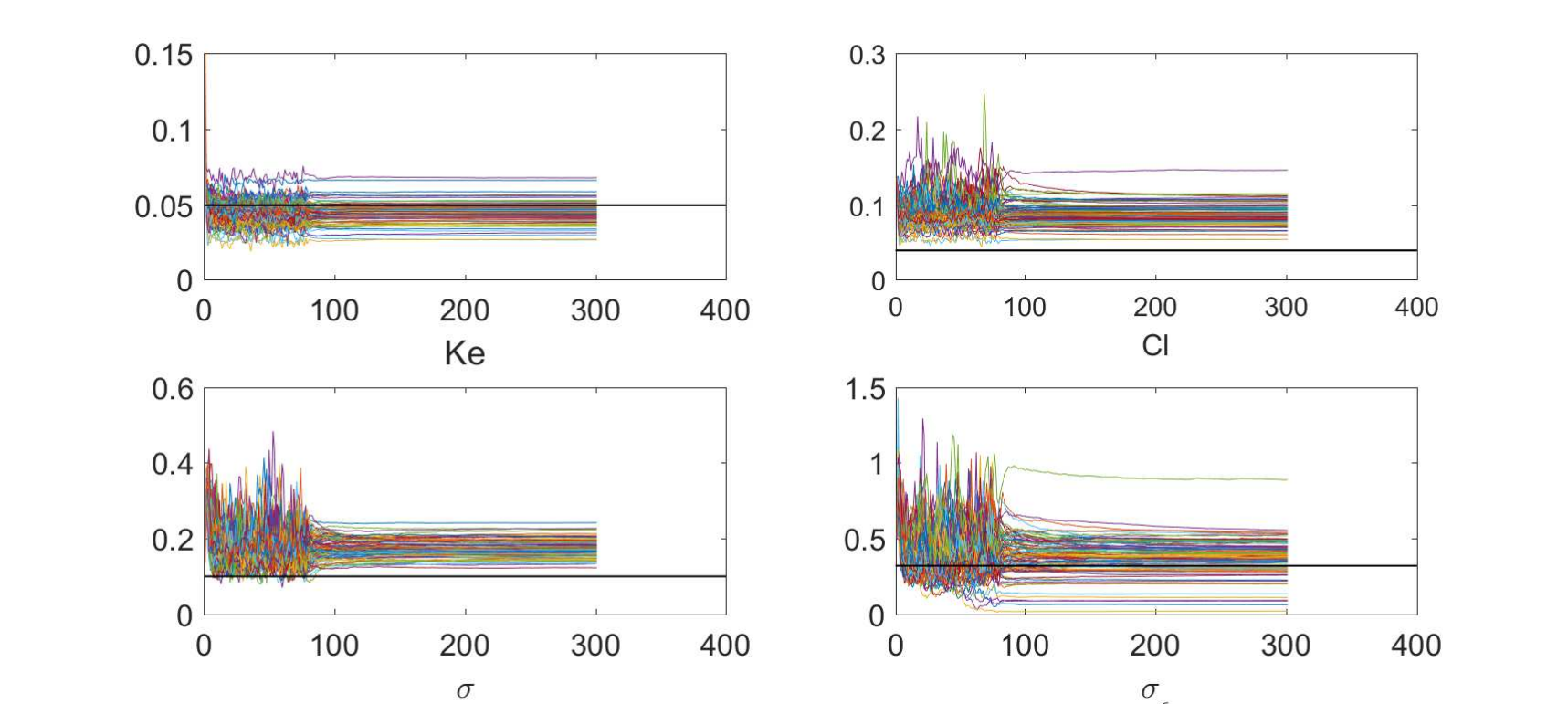} 
\caption{\footnotesize{Theophylline model: $K=100$ iterations of 100 repetitions of SAEM-SMC using $M=500$. Top: $K_e$ (left) and $Cl$ (right). Bottom: $\sigma$ (left) and $\sigma_{\epsilon}$ (right). Horizontal lines are the true parameter values.}}
\label{fig:theophylline-SAEM-SMC-iter}
\end{figure}

 Same as in section \ref{sec:nonlingauss}, for SAEM-SMC we run a number of independent repetitions of the estimation procedure: the dataset is shorter than in section \ref{sec:nonlingauss} and despite the need to resort to numerical integration of the SDE, we are able to run 100 estimation procedures in about 300 seconds overall. Each repetition generates a different dataset  using the true parameter values, then for each repetition SAEM-SMC is initialized at the same  parameter values $K_e=0.15$, $Cl=0.135$, $\sigma=0.135$ and $\sigma_{\varepsilon}=0.502$. We always use a warmup of $K_1=80$ iterations, $K=300$, $M=500$ particles and $\bar{M}=100$. We observed an $ESS>100$ at the last time point for each simulation. See Table \ref{tab:theophylline} and Figure \ref{fig:theophylline-SAEM-SMC-iter} for results: clearly $Cl$ and $\sigma$ are not identified. However these results can be improved, at least for $Cl$: for state-space models having additive Gaussian noise and an SDE model discretised using Euler-Maruyama, \cite{golightly2011bayesian} propose a SMC filter where forward simulation of the particles is not blind to data (unlike the bootstrap filter). We refer the reader to \cite{golightly2011bayesian} for details and report results using their approach as SAEM-GW in Table \ref{tab:theophylline}. While $Cl$ is very well identified, the system noise $\sigma$ is still elusive.
 
With SAEM-SL we only need to set the vector of summaries $(\bS(\bx_r^*),\bS(\by_r^*))$. The vector $\bS(\bx_r^*)$ contains (i) the median values of $\bX_{0:N}^*$ ; (ii) the median absolute deviation of $\bX_{0:N}^*$, (iii) a statistic for $\sigma$  computed from $\bX_{0:N}^*$  (see below) and (iv) $(\sum_j(Y_{j}^*-X_{j}^*)^2/n)^{1/2}$ with $X_{j}$ the $j$th element of the $n$ interpolated values $\bX_{1:n}$. Vector $\bS(\by_r^*)$ contains: (i) the median value of $\by_r^*$; (ii) its median absolute deviation; (iii) the slope of the line connecting the first and last simulated observation $(Y_n^*-Y_1^*)/(t_n-t_1)$, since concentrations show a markedly decaying behaviour. 
In \cite{miao} it is given that, for an SDE of the type $dX_t=\mu(X_t)dt+\sigma g(X_t)dW_t$ with $t\in [0,T]$, we have 
\[
\frac{\sum_{\Gamma}|X_{i+1}-X_i|^2}{\sum_{\Gamma}g(X_i)(t_{i+1}-t_i)}\rightarrow \sigma^2 \qquad as\quad |\Gamma|\rightarrow 0
\]
where the convergence is in probability and $\Gamma$ a partition of $[0,T]$.
Therefore we deduce that using the discretization $\{X_0,X_1,...,X_N\}$  produced by the Euler-Maruyama scheme, we can take the square root of the left hand side in the limit above, which should be informative for $\sigma$. We use this as the third summary statistic in $\bS(\bx_r^*)$.

\begin{figure}
\centering
\includegraphics[width=16cm,height=7cm]{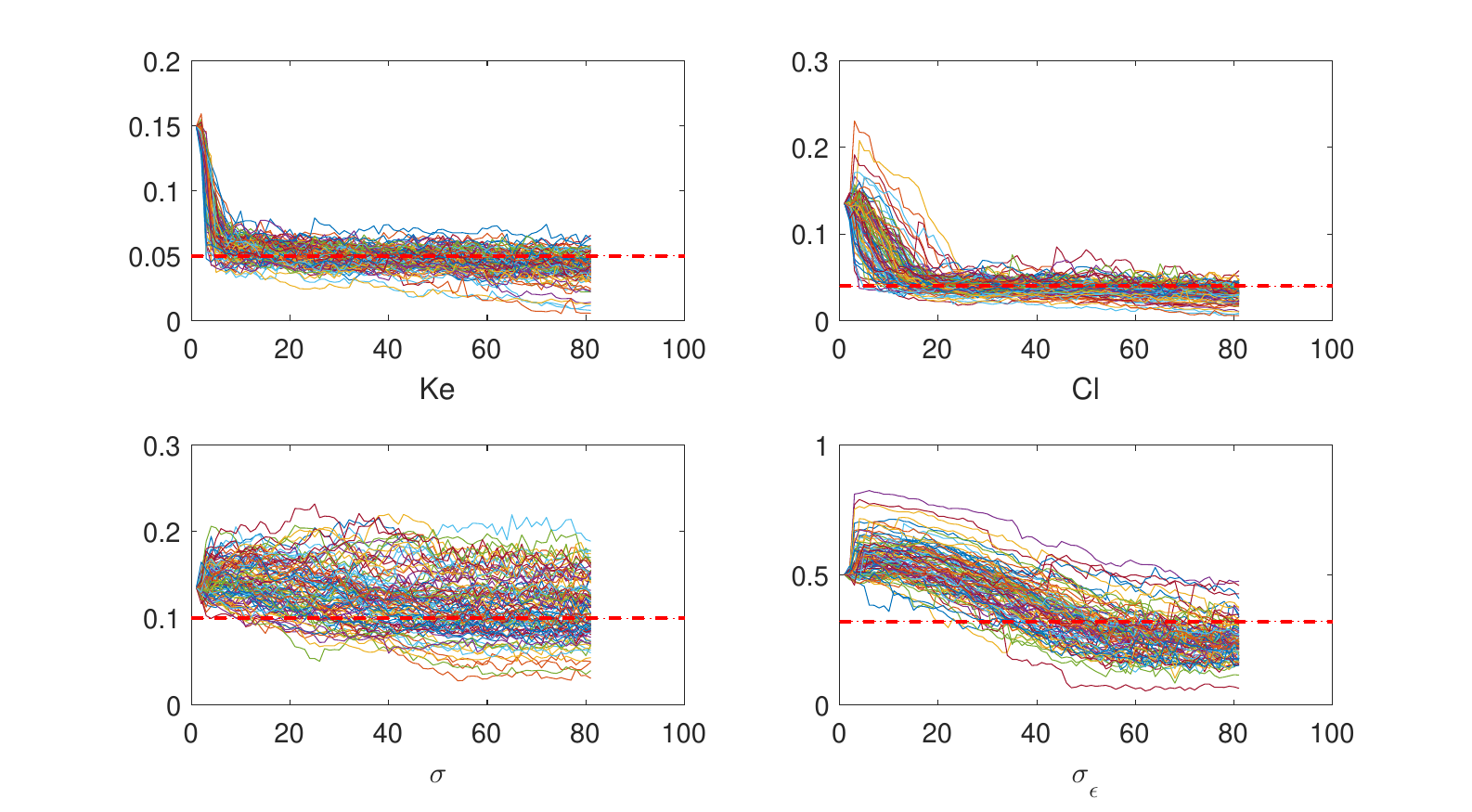} 
\caption{\footnotesize{Theophylline model: $K=80$ iterations of 100 repetitions of SAEM-SL using $R=200$. Top: $K_e$ (left) and $Cl$ (right). Bottom: $\sigma$ (left) and $\sigma_{\epsilon}$ (right). Dashed lines are the true parameter values.}}
\label{fig:theophylline-SAEM-SL-iter}
\end{figure}

\begin{figure}
\centering
\includegraphics[width=14cm,height=6cm]{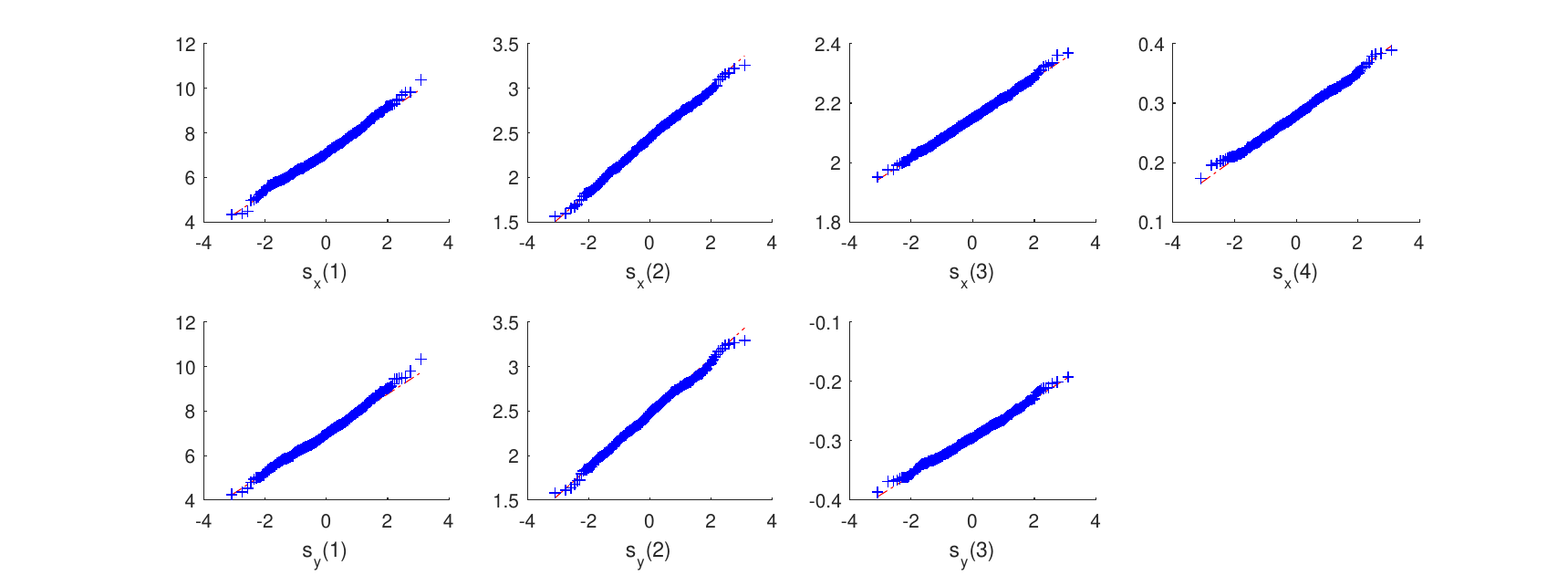}
\caption{\footnotesize{Theophylline model: normal qq-plots for $\bS(\bx_r)$ and $\bS(\by_r)$ corresponding to the last iteration of SAEM-SL.}}
\label{fig:theophylline-qqplots}
\end{figure}

We used SAEM-SL on the same simulated data produced when implementing SAEM-SMC. We considered  $R=200$ simulated summaries and, since for this example SAEM-SL is computationally more intense than SAEM-SMC, we consider $K_1=50$ and $K=80$, with $L=30$ for the number of iterations in the maximization step. Notice for this example we found benefit in using robust methods for the computation of sample means and covariances, downweighting summaries falling in the tails of the multivariate Gaussian synthetic likelihood. Specifically, here we compute the moments \eqref{eq:synthetic-sample-moments} using the method in \cite{olive2010robust}.
See Table \ref{tab:theophylline} and Figure \ref{fig:theophylline-SAEM-SL-iter} for results. Notice that simulations for SAEM-SMC and SAEM-SL start at the same parameter values, even though from Figures \ref{fig:theophylline-SAEM-SMC-iter}--\ref{fig:theophylline-SAEM-SL-iter} it may seem otherwise (that is because SAEM-SMC reaches almost immediately the final values while SAEM-SL converges more slowly). 
 SAEM-SL produces satisfactory results on all parameters. 
For  one of the one-hundred repetitions, Figure \ref{fig:theophylline-qqplots} shows the normal qq-plots for the seven summary statistics (the four statistics in $\bS(\bx_r)$ and the three in $\bS(\by_r)$), generated at the optimum returned by SAEM-SL. Also for this example, there are no major departures from normality. 

\begin{figure}
\centering
\includegraphics[width=12cm,height=8cm]{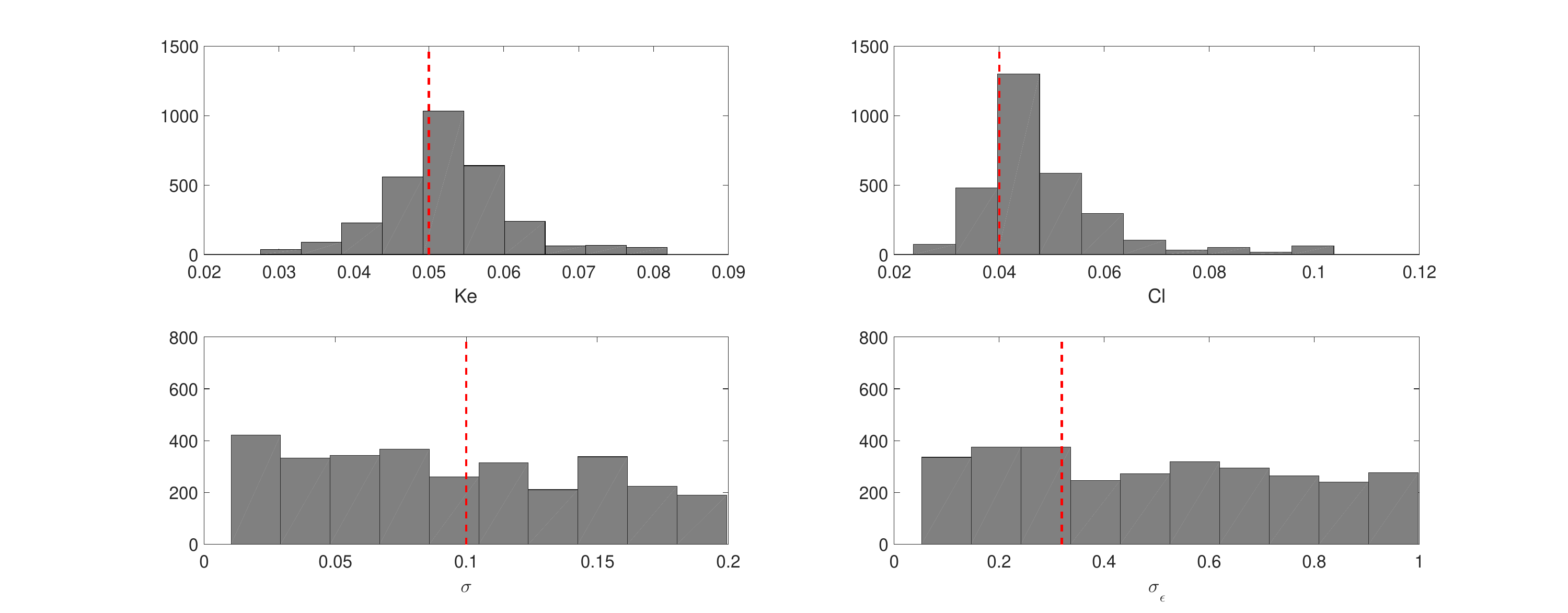}
\caption{\footnotesize{Theophylline model: marginal posteriors from the pseudo-marginal Bayesian SL using $R=2,000$. Vertical lines are true parameter values.}}
\label{fig:theophylline-bayesSL-hist}
\end{figure}

We now run a single instance of the pseudo-marginal Bayesian SL algorithm of \cite{price2016bayesian}. We impose independent uniform  priors $K_e\sim U(0.01,1)$, $Cl\sim U(0.01,20)$, $\sigma\sim U(0.01,0.2)$ and $\sigma_\varepsilon\sim U(0.05,1)$ and run 5000 MCMC iterations. Parameters were proposed using the adaptive Gaussian random walk of \cite{haario-saksman-tamminen}, obtaining an acceptance rate of about 25\%.  We first consider $R=200$, same as for SAEM-SL. Posterior means and 95\% posterior intervals for each parameter are: $\hat{K}_e=0.052$ [0.027,0.074], $\hat{Cl}=0.048$ [0.027,0.091], $\hat{\sigma}=0.105$ [0.024,0.195], $\hat{\sigma}_\varepsilon=0.541$ [0.087,0.9969]. 
We notice the first two parameters are correctly identified while the latter two parameters are essentially unidentified. This can also be noticed from their MCMC trace plots, spanning the support of the corresponding priors (plots not reported for brevity). Results can be partially improved using $R=2,000$, producing better identification for the first two parameters but not for the latter two, see Figure  \ref{fig:theophylline-bayesSL-hist}. Therefore, using uniform priors as suggested in \cite{wood2010statistical} when the MAP is the only object of interest is not appropriate here and strongly informative priors for $(\sigma,\sigma_\varepsilon)$ might be needed.

\begin{table}
\centering 
\begin{tabular}{lllll}
\hline
 & $K_e$  &  $Cl$ & $\sigma$&  $\sigma_{\varepsilon}$\\ 
\hline
true values & 0.050& 0.040&   0.100& 0.319\\
SAEM-SMC&  0.045  [0.042,0.049]& 0.085 [0.078,0.094] &   0.171 [0.158,0.184] & 0.395 [0.329,0.465] \\
SAEM-GW & 0.053 [0.049,0.058] & 0.039 [0.035,0.043] & 0.704 [0.549,0.963] & 0.175 [0.119,0.304]\\
SAEM-SL & 0.045 [0.037,0.049]  &  0.032 [0.027,0.038]  &  0.113 [0.088,0.144]  &  0.241 [0.200,0.294]\\
\hline
\end{tabular}     
\caption{\footnotesize{Theophylline: medians and $1^{st}-3^{rd}$ quartiles for estimates obtained on 100 independent simulations using SAEM-SMC, SAEM-GW and SAEM-SL.}}
\label{tab:theophylline}
\end{table}

\subsection{Lotka-Volterra model}\label{sec:lv}

The Lotka–Volterra model (LV) is a stochastic Markov jump process that describes the continuous time
evolution of a population of prey ($X_1$) interacting with a population of predators ($X_2$). The populations are subject to three possible
reactions: (a) reproduction, (b) predator-prey interaction (consumption of prey by predator, in turn influencing predator reproduction rate), (c) death of predators due to natural causes. 
These reactions occur at random times and depend on unknown rates $(c_1,c_2,c_3)$ that influence the amount of individuals in the two species, for given initial population sizes $x_{10}$ and $x_{20}$. Realizations for the LV model can be simulated exactly using the so-called ``Gillespie algorithm'' \citep{gillespie1977exact}. We set $x_{10}=x_{20}=100$ and $(c_1,c_2,c_3)=(0.5,0.0025,0.3)$ as in \cite{fearnhead-prangle(2011)}. 

\begin{figure}
\centering
\includegraphics[width = 15cm,height=8cm]{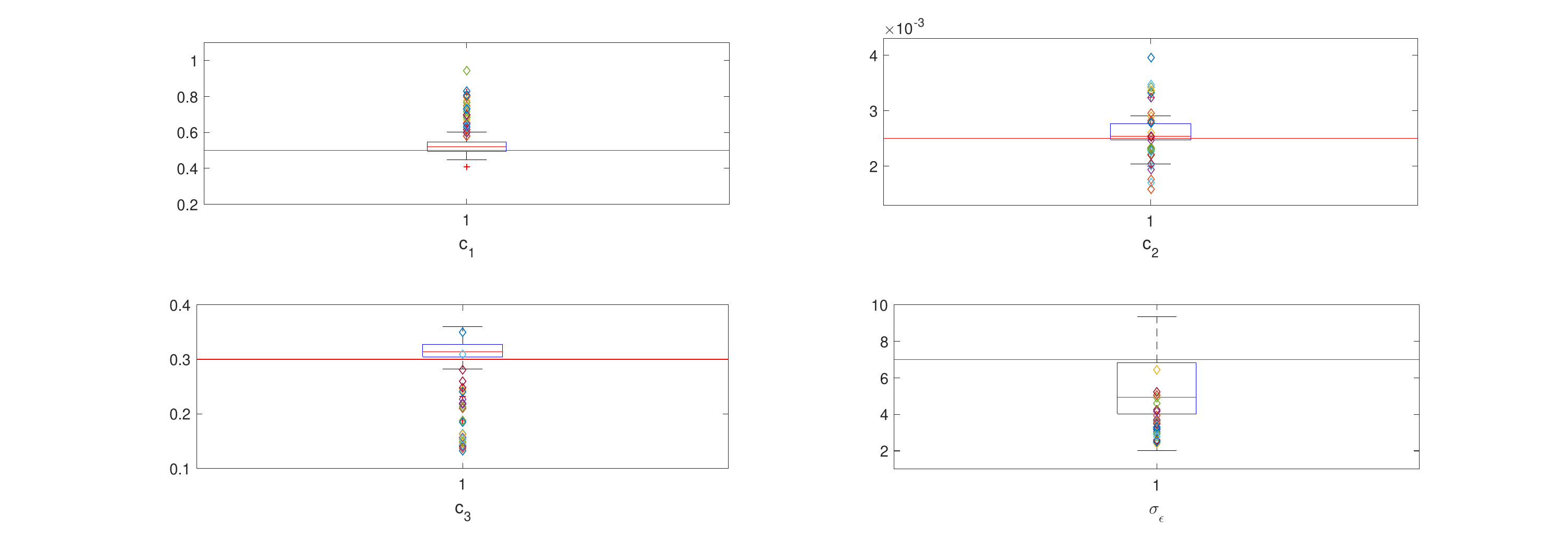}
\caption{\footnotesize{Lotka--Volterra: boxplots of the SAEM-SL estimates using $R=1,000$. Parameter starting values are denoted with $\diamondsuit$. Horizontal lines are true parameter values.}}
\label{fig:lv-boxplot}
\end{figure}

\begin{figure}
\centering
\includegraphics[width = 15cm,height=8cm]{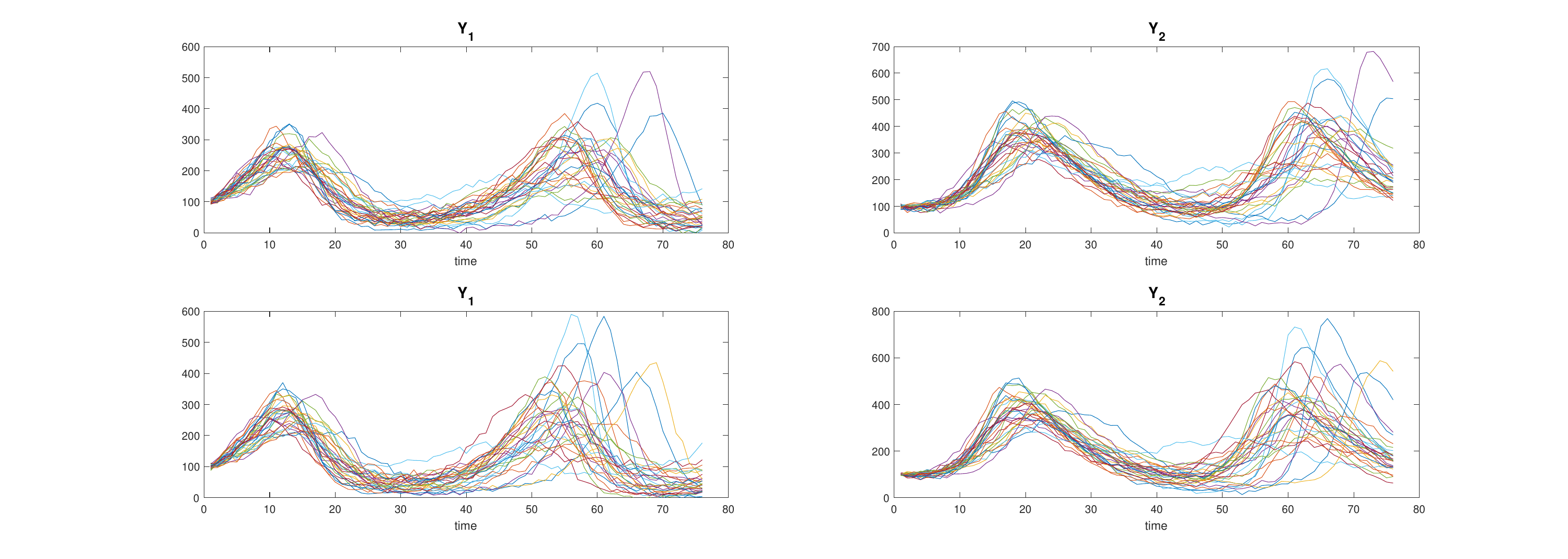}
\caption{\footnotesize{Lotka--Volterra: (top) thirty realizations of LV (left $Y_1$, right $Y_2$) using the ground truth parameters. (bottom) Thirty realizations of LV using the medians of the thirty parameter estimates obtained with SAEM-SL. We used the same seed for pseudo-random numbers to produce plots on top and bottom.}}
\label{fig:lv-trajectories}
\end{figure}

In our experiment each simulation took place for a total of 30 time units. We recorded the values of $X_1$ and $X_2$
after every 0.4 time units, resulting in two time series $\{x_{1,t},x_{2,t}\}_{t=1:T}$ of 76 values each. Finally we added independent realizations of homoscedastic Gaussian noise $\mathcal{N}(0,\sigma_{\varepsilon}^2)$ to each of the recorded realizations to obtain data measurements $\{y_{1,t},y_{2,t}\}_{t=1:T}$ from variables $(Y_1,Y_2)$ with $\sigma_\varepsilon=7$ and $T=76$. We kept the initial states $x_{10}=x_{20}$ fixed to their true values and estimate $\btheta=(c_1,c_2,c_3,\sigma_\varepsilon)$ with SAEM-SL. We denote with $\bx_r=(\bx_{1,r},\bx_{2,r})$ the simulated $T\times 2$ matrix of stochastic realizations for $(X_1,X_2)$ and with $\by_r=(\by_{1,r},\by_{2,r})$ the corresponding noisy versions obtained after adding Gaussian noise.  We first formulate the following summary statistics (subject to amendment as we explain below): for $\tilde{\bS}(\bx_r)$ we consider (i) sample means of $\bx_{1,r}$ and of $\bx_{2,r}$; (ii) log-variances of $\bx_{1,r}$ and $\bx_{2,r}$; (iii) lag-one autocorrelation $\rho_1(x)$ and $\rho_2(x)$ for $\bx_{1,r}$ and $\bx_{2,r}$ respectively; (iv) cross-correlation $\rho_{12}(x)$ between $\bx_{1,r}$ and $\bx_{2,r}$. For $\tilde{\bS}(\by_r)$ we consider the analogous statistics as for $\tilde{\bS}_(\bx_r)$. Intuitively, correlations and autocorrelations have very asymmetric distributions, and our initial inference attempts with $(\tilde{\bS}(\bx_r),\tilde{\bS}(\by_r))$ were failures (results not reported). However, in this case it was easy to enforce approximate Gaussianity by applying Box-Cox transformations to these preliminary summaries, and the resulting summaries $(\bS(\bx_r),\bS(\by_r))$ were used to produce reported results. Hence $\bS(\bx_r)$ is the same as $\tilde{\bS}(\bx_r)$ except for the lag-one autocorrelations $(\rho_1(x))^{82}$ and $(\rho_2(x))^{59.4}$ and cross-correlation $(\rho_{12}(x)+1)^{0.1}$. Similarly, $\bS(\by_r)$ is as $\tilde{\bS}(\by_r)$ but with $(\rho_1(y))^{63.6}$, $(\rho_2(y))^{61}$ and $(\rho_{12}(y)+1)^{-0.13}$. We produced thirty independent noisy datasets of $(Y_1,Y_2)$ using the same ground-truth parameter values, then used SAEM-SL with $R=1,000$. The starting parameter values were randomly drawn from a multivariate Gaussian centred at $(0.7,0.001,0.1,3)$, see Figure \ref{fig:lv-boxplot} to notice the spread of the starting values marked with diamonds $\diamondsuit$. Clearly the reaction rates are well estimated, while $\sigma_\varepsilon$ is underestimated.
Notice that for certain carefully tuned values of $\btheta$ the
two species exhibit an oscillatory behaviour, typical of natural ecological systems. Our ground truth values for $(c_1,c_2,c_3)$ have been chosen to give rise to oscillatory behaviour. However, as remarked in \cite{papamakarios2016fast}, only a small subset of parameters give rise to such oscillatory behaviour, hence in a Bayesian framework the parameter posteriors are narrow and expected to be tightly peaked around the true parameter values. However Figure \ref{fig:lv-trajectories} shows that we recovered the true dynamics correctly.

\section{Summary}\label{sec:summary}

We have introduced a new method for approximate maximum likelihood estimation of the parameters of intractable models. Under this framework, our method is able to deal with a large class of modelling scenarios, and both ``static'' (example in the Supplementary Material) and ``dynamic'' models (examples in sections \ref{sec:nonlingauss}--\ref{sec:lv}) can be accommodated. We started by illustrating the stochastic approximation EM algorithm (SAEM, \citealp{Delyon1999}) as one of the possible ways to implement an EM algorithm. To fully exploit the computational benefits of SAEM, namely convergence to a (local) maximizer of the data likelihood, it is required to analytically compute the complete likelihood of the model and derive the corresponding sufficient statistics. The latter step is far from being trivial (if at all possible) for most models of realistic complexity. Our SAEM-SL method makes use of the synthetic likelihoods (SL) approach proposed in \cite{wood2010statistical}: SL requires from the modeller the specification of ``appropriate'' (informative) summary statistics encoding the information about the parameter $\btheta$ that is contained in the available data. These summaries are assumed to follow a Gaussian distribution and we find that this assumption is convenient for exploitation in a SAEM context, as Gaussian likelihoods have trivial to compute sufficient statistics, which we obtain from SL simulations. Our approach constructs a version of SL for the ``complete synthetic loglikelihood'' and plugs it within SAEM. As a result, it bypasses the analytic calculation of the complete likelihood of the model by introducing a Gaussian approximation. SAEM-SL results in a plug-and-play, likelihood-free, approximated version of SAEM. 
Under ideal scenarios, where the user-specified summaries are sufficient statistics for $\btheta$ \textit{and} are also Gaussian distributed, then SAEM-SL is equivalent to the standard SAEM and therefore should return a  stationary point of the true data-likelihood. 

In four simulation studies (one is available in the Supplementary Material) we have shown the good performance of the method, which requires minimal tuning. 
However SAEM-SL requires from the modeller a set of summary statistics: this operation is clearly subjective and delicate. A possibility to automatize the process of selection of the statistics is to run a semi-automatic summaries selection algorithm as described in \cite{fearnhead-prangle(2011)}, from within an approximate Bayesian computation framework, then plug the constructed summaries into SAEM-SL. We have not considered the possibility to use the semi-automatic selection approach in the present work, and a study of the implications is left for future research. In conclusion SAEM-SL is an appealing likelihood-free version of SAEM for intractable models. We have performed several comparisons with well established methodologies (such as iterated filtering, particle marginal methods, approximate Bayesian computation and SAEM incorporating a sequential Monte Carlo step) and while SAEM-SL performs satisfactorily against available alternatives, in challenging settings the ``best'' approach for a specific problem is often a compromise between computational feasibility and statistical efficiency.

\section*{Acknowledgements} 
The work was partially supported by the Swedish Research Council under grant 2013-5167.
This paper is published in \textit{Communications in Statistics - Simulation and Computation}, \url{https://doi.org/10.1080/03610918.2017.1401082}.

We thank Christopher Drovandi (Queensland University of Technology) for valuable discussion and for suggesting the Cholesky factorization check when sampling from a multivariate Gaussian.

\bibliographystyle{Chicago}  
\bibliography{biblio}

\section*{Supplementary material}

This section contains the following items:

\begin{description}

\item[Inference for g-and-k distributions] A simulation study has been conducted to show the performance of SAEM-SL for a ``static'' model, where observations arise from a $g$-and-$k$ distribution corrupted with noise. A comparison with an approximate Bayesian computation (ABC) MCMC algorithm is also performed. 

\item[Sufficient statistics for the example in section \ref{sec:theoph}] These statistics are necessary to run the standard SAEM algorithm, but are not necessary to use SAEM-SL. 

\item[MATLAB package for the first and second example] MATLAB files to run SAEM-SL for the examples in section \ref{sec:nonlingauss}-\ref{sec:theoph} are available at \url{https://github.com/umbertopicchini/SAEM-SL}.

\end{description}

\section{A static model: noisy data from a $g$-and-$k$ distribution}\label{sec:g-and-k}

We now consider a ``static'' model, namely a $g$-and-$k$ distribution corrupted with noise. Noise-free versions of samples from  $g$-and-$k$ distributions have been considered numerous times in the ABC literature (e.g. \citealp{allingham2009bayesian,fearnhead-prangle(2011),picchini2016approximate}). This is a flexibly shaped distribution that is used to model non-standard data
through a small number of parameters. It is defined by its inverse distribution
function, but has no closed form density hence it is an example of model with an intractable likelihood. Therefore it cannot be dealt with using, say, standard SAEM methods, as the explicit computation of the complete likelihood (and its sufficient statistics) is impossible. However it is trivial to sample from a $g$-and-$k$ distribution and therefore ABC is an appealing methodology for this problem. The quantile function (inverse distribution function) is given by 
\begin{equation}
F^{-1}(z;A,B,c,g,k)= A+B\biggl[1+c\frac{1-\exp(-g\cdot r(z))}{1+\exp(-g\cdot r(z))}\biggr](1+r^2(z))^kr(z)
\label{eq:g-k-inverse}
\end{equation}
where $r(z)$ is the $z$th standard normal quantile, $A$ and $B$ are location and scale parameters and $g$ and $k$ are related to skewness and kurtosis. Parameters restrictions are $B>0$ and $k>-0.5$.
An evaluation of \eqref{eq:g-k-inverse} returns a draw ($z$th quantile) from the $g$-and-$k$ distribution or, in other words, the $j$th sample $r_j:=r_j(z)\sim \mathcal{N}(0,1)$ produces a draw $x_j:=F^{-1}(\cdot;A,B,c,g,k)$ from the $g$-and-$k$ distribution. However, unlike in previously mentioned references, we consider as data the vector $\by=(y_1,...,y_n)$, where $y_j=x_j+\varepsilon_i$, with i.i.d. noise $\varepsilon_j\sim \mathcal{N}(0,\sigma^2_\varepsilon)$, where the $\varepsilon_j$'s are independent of the $x_j$'s, $j=1,...,n$. Also, denote $\bx=(x_1,...,x_n)$. Notice that because SAEM-SL is an EM-type algorithm, and therefore it is suitable for ``incomplete data'', we would not be able to apply SAEM-SL to data observed directly as realizations from \eqref{eq:g-k-inverse}. That is while ABC methods can in principle accommodate inference based on either noisy data $\by$ and noise-free data $\bx$, SAEM-SL can only deal with the former. We found the parameter $g$ to be of difficult identification and in the following we keep it fixed at its true value (see below): hence we assume $\btheta=(A,B,k,\sigma_\varepsilon)$ as parameter of interest, by noting that it is customary to keep $c$ fixed to $c=0.8$ (\citealp{drovandi2011likelihood,rayner2002numerical}).

 We initially consider the summaries $\tilde{\bS}(\bx)=(S_{A,\bx},S_{B,\bx},S_{g,\bx},S_{k,\bx},P_{20,\bx},P_{30,\bx},P_{70,\bx},P_{80,\bx})$, where $P_{q,\bx}$ is the $q$th empirical percentile of $\bx$, whereas the remaining summaries are from \cite{drovandi2011likelihood}:
\begin{align*}
S_{A,\bx}&=P_{50,\bx} & S_{B,x}&=P_{75,\bx}-P_{25,\bx},\\ 
S_{g,\bx}&=(P_{75,\bx}+P_{25,\bx}-2S_{A,\bx})/S_{B,\bx} & S_{k,\bx}&=(P_{87.5,\bx}-P_{62.5,\bx}+P_{37.5,\bx}-P_{12.5,\bx})/S_{B,\bx}.
\end{align*}
That is $S_{A,\bx}$ and $S_{B,\bx}$ are the median and the inter-quartile range of $\bx$ respectively.
We define summaries $\tilde{\bS}(\by)$ for observed data in the analogous way as for $\bx$, that is by plugging $\by$ in place of $\bx$ in the summaries above. However we found that working with $\tilde{\bS}(\bx)$ and $\tilde{\bS}(\by)$ produces unsatisfactory results, because the distributions of some of the simulated summaries are markedly asymmetric, i.e. far from being even approximately Gaussian. Therefore in practice we work with $\bS(\by):=\log(\tilde{\bS}(\by)+\nu)$ and $\bS(\bx):=\log(\tilde{\bS}(\bx)+\nu)$, where $\nu>0$ is a constant set so that the argument of the logarithms is strictly positive, and of course the same $\nu$ has to be used for $\bS(\bx)$ and $\bS(\by)$ during the execution of SAEM-SL.
Therefore SAEM-SL is implemented with $\bS=(\bS(\by),\bS(\bx))$.  For the specific data $\by$ simulated with the setting given below, $\nu=50$ was found to be appropriate.

\begin{figure}
\centering
\includegraphics[width=17cm,height=8cm]{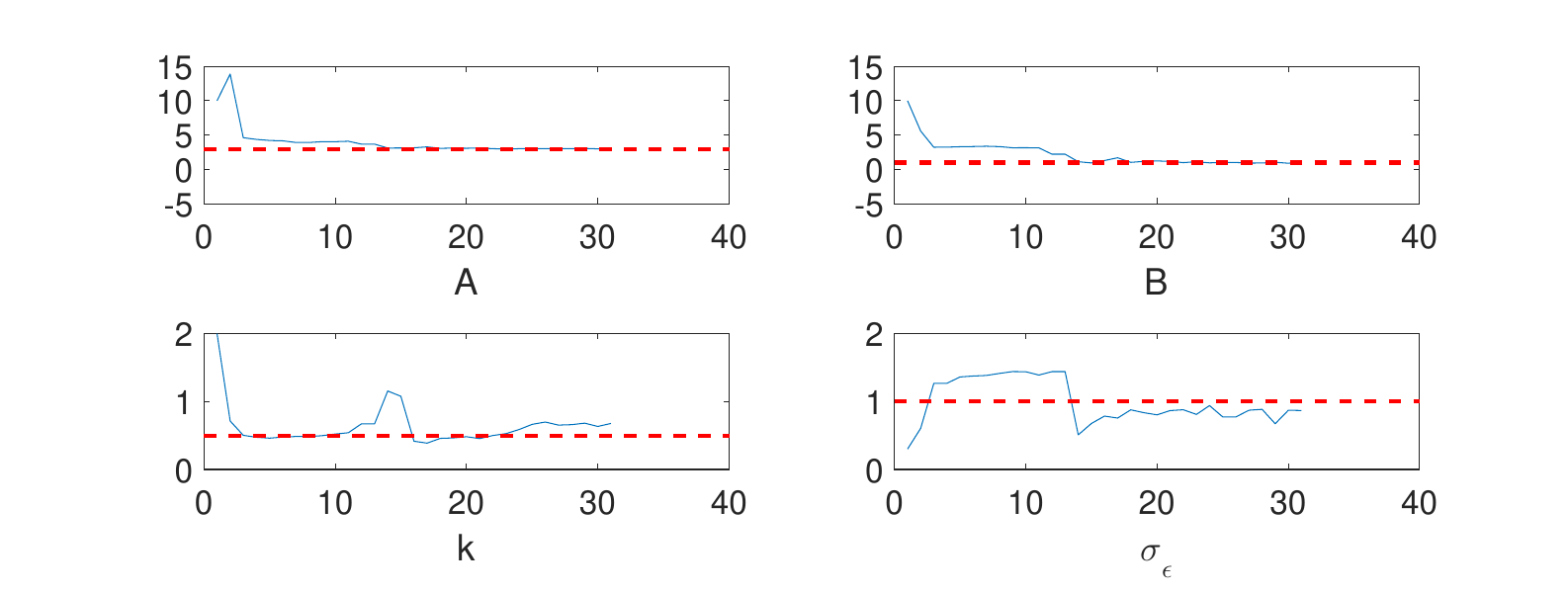}
\caption{\footnotesize{g-and-k distribution: evolution of SAEM-SL. Dashed lines are the true parameter values.}}\label{fig:gk_saem_trajectories}
\end{figure}

Here we intend to compare SAEM-SL with an ABC algorithm. Therefore we produce a single dataset having length $n=500$, generated with $\btheta=(A,B,k,\sigma_\varepsilon)=(3,1,0.5,1)$ (we keep $c=0.8$ and $g=2$ fixed). 
Starting values for SAEM-SL are $A=10$,          $B=10$,       $k= 4$ and      $\sigma_\varepsilon=0.3$. We run SAEM-SL with $R=3,000$, $K_1=10$ and $K=20$ and use $L=40$ iterations for the M-step. The result is in Figure \ref{fig:gk_saem_trajectories}. The simulation is relatively computer intensive, as computing the summaries (hence the percentiles) requires sorting procedures on each of the $R$ simulated data. Our SAEM-SL estimation required about 10 minutes of computation.

\subsubsection*{Bayesian estimation via ABC-MCMC}\label{sec:abc-mcmc}

Here we consider a comparison with the ``gold standard'' methodology for intractable likelihoods, that is approximate Bayesian computation (ABC). Several possible ABC methods could be considered: we choose an ABC-MCMC sampler, essentially a trivial modification of the one proposed in \cite{marjoram2003markov}, see for example \cite{sisson-fan(2011)}. As  shown in e.g. \cite{picchini2016approximate} it is possible to estimate parameters of  noise-free data from $g$-and-$k$ distributions using ABC-MCMC, and we now consider the case of noisy data. Briefly, with ABC the goal is to sample from an approximate posterior $\pi_{\delta}(\btheta,\bz|\by)$ defined on the space of $\btheta$ augmented with the space of $\bz$. Here $\bz$ denotes synthetic observations defined on the same space as the actual observations $\by$, that is if $\by\in\mathcal{Y}$ are noisy observations then so are the $\bz\in\mathcal{Y}$, and $\bz$ should be simulated with the same generating model assumed for $\by$. However, typically in ABC studies a set of summary statistics is introduced to break the curse-of-dimensionality, and the resulting posterior is (by disregarding normalizing factors) 
\begin{equation}
\pi_{\delta}(\btheta,\bz|\rho(\bS(\by),\bS(\bz))\propto J_{\delta}(\bS(\by),\bS(\bz)){p(\bz|\btheta)\pi(\btheta)},
\label{eq:abc-posterior}
\end{equation}
with $p(\bz|\btheta)$ the likelihood function based on $\bz$ and $\pi(\theta)$ the prior for $\btheta$.
Here $\delta>0$ is a threshold value and $J_{\delta}(\bS(\by),\bS(\bz))$ is a positive function assigning larger weights to values of $\btheta$ such that $\rho(\bS(\by),\bS(\bz))<\delta$ for some appropriate distance $\rho(\cdot,\cdot)$. It can be shown that for a small enough $\delta$ the marginal ABC posterior $\pi_{\delta}(\btheta|\bS(\by))=\int \pi_{\delta}(\btheta,\bz|\rho(\bS(\by),\bS(\bz))) d\bz $ is ``close'' to the true marginal $\pi(\btheta|\bS(\by))$, if the summary statistics are informative for $\btheta$. Essentially, an ABC-MCMC algorithm produces a Markov chain for $\btheta$ having stationary distribution $\pi_{\delta}(\btheta|\bS(\by))$. It should be remarked that in ABC $\bS(\by)$ and $\bS(\bz)$ are the same set of summary functions, only applied to different arguments, as $\bz$ and $\by$ are assumed to be defined on the same space and generated with the same underlying mechanism. For SAEM-SL the summaries we denoted with $\bS(\by)$ and $\bS(\bx)$ in general \textit{do not} have to be the same functions, as $\bx$ is a noise-free version of $\by$ hence these are defined on different spaces; however for this example we chose $\bS(\by)$ and $\bS(\bx)$ to be the same set of functions.

To implement ABC-MCMC we choose a Gaussian kernel for $J_{\delta}(\bS(\by),\bS(\bz))$, given by
\[
J_{\delta}(\bS(\by),\bS(\bz))\propto \exp\{-(\bS(\bz)-\bS(\by))'\mathbf{\Omega}^{-1}(\bS(\bz)-\bS(\by))/2\delta^2\}
\] 
where $'$ denotes transposition and $\mathbf{\Omega}$ is a positive definite matrix. For simplicity we assume a diagonal $\mathbf{\Omega}$ with elements $\mathbf{\Omega}=\mathrm{diag}\{\omega^{2}_1,...,\omega^{2}_{d_s}\}$, with $d_s=\dim \bS(\by)=\dim \bS(\bz)$. When the elements in vector $\bS(\by)$ are  varying approximately on the same range of values it is possible to consider $(\omega^{2}_1,...,\omega^{2}_{d_s})=(1,...,1)$, however in general the variability of the statistics is unknown and, depending on the type of data and the underlying model, these can have very different magnitude. 
The interested reader is referred to section 3.1 in \cite{picchini2016approximate} for further details (and disregarding the ``data cloning'' approach there exposed). 

For $\bS(\by)$ and $\bS(\bz)$ we consider the same set of summaries used with SAEM-SL and the same starting values for the parameters. 
We run two attempts of an ABC-MCMC algorithm, with independent uniform priors $U(0,1)$ for $A$, $B$ and $k$ while we set $\sigma_\varepsilon\sim \Gamma(2,1)$, that is a Gamma distribution with mean 2. Parameters were proposed using an adaptive Metropolis algorithm with Gaussian innovations \citep{haario-saksman-tamminen}. At the first (pilot) attempt we use $(\omega^{2}_1,...,\omega^{2}_{d_s})=(1,...,1)$,  
and let $\delta$ decrease every 20,000 iterations in $\delta\in\{0.03, 0.007, 0.003\}$, for a total of 60,000 iterations, where the $\delta$'s were chosen to target an acceptance rate of 1--3\% at the smallest $\delta$, usually considered a good compromise between accuracy and computational budget. Results were not encouraging, because the summaries vary on different scales but we assigned unit weight to each of them. However, we also collect the 20,000 summary statistics simulated at the smallest $\delta$, i.e. at $\delta=0.003$ and from these statistics we compute the median absolute deviation MAD for each coordinate of the accepted $\bS(\bz)$ and define $(\omega_1,...,\omega_{d_s}):=(\mathrm{MAD}_1,...,\mathrm{MAD}_{d_s})$. We plug the new weights into $\mathbf{\Omega}$ for a further run of ABC-MCMC, this time using $\delta\in\{8, 3, 1, 0.3\}$, and the $\delta$'s had to be modified as a consequence of the different weights introduced, again targeting an acceptance rate of 1--3\% at the smallest $\delta$.
We use the parameter draws simulated in correspondence of $\delta=0.3$ to calculate the parameters posterior means and $95\%$ posterior intervals, and these result in: $\hat{A}=3.03$ [2.71,3.36], $\hat{B}=1.32$ [0.48,2.36], $k=0.51$ [0.05,1.45], $\sigma_\varepsilon=0.90$ [0.49,1.38]. 

Here the strength of ABC methods is on full display: ABC is not constrained by any parametric assumption regarding the distribution of the summaries, and when these are informative ABC is probably the go-to choice. 
The essence of the comparison is that tuning ABC algorithms is not trivial. However, for each iteration of ABC-MCMC we only need to simulate a single realization of $\bz$, while for each iteration of SAEM-SL we need at least $L\times R$ simulations from the model. However, a proper comparison between SAEM-SL and ABC is not problem independent. For example, in stochastic dynamical modelling an ABC-MCMC sampler will seldom produce accurate results and an ABC-SMC approach will usually be preferred (see e.g. \citealp{toni2009approximate}), this increasing the computational effort considerably.

\section{Theophylline example: sufficient statistics for SAEM}\label{sec:suffstats}
Recall that the statistics we are about to construct are required for the standard SAEM to run (e.g. SAEM-SMC) but not for SAEM-SL.
The complete likelihood is given by
\[
 p(\bY,\bX_{0:N};\btheta)=p(\bY|\bX_{0:N};\btheta)p(\bX_{0:N};\btheta)=\prod_{j=1}^n p(Y_j|X_j;\btheta)\prod_{i=1}^Np(X_i|X_{i-1};\btheta)
\]
where the unconditional density $p(x_0)$ is disregarded in the last product since we assume $X_0$ deterministic. Hence the complete-data loglikelihood is
\[
L_c(\bY,\bX_{0:N};\btheta) =\sum_{j=1}^n \log p(Y_j|X_j;\btheta)+\sum_{i=1}^N \log p(X_i|X_{i-1};\btheta).
\]
Here $p(y_j|x_j;\btheta)$ is a Gaussian density with mean $x_j$ and variance $\sigma^2_{\varepsilon}$. The transition density $p(x_i|x_{i-1};\theta)$ is not known for this problem, hence we approximate it with the Gaussian density induced by the Euler-Maruyama scheme, that is
\[
p(x_i|x_{i-1};\btheta)\approx \frac{1}{\sigma\sqrt{2\pi x_{i-1}h}}\exp\biggl\{-\frac{\bigl[x_i-x_{i-1}-(\frac{Dose\cdot K_a \cdot K_e}{Cl}e^{-K_a \tau_{i-1}}-K_ex_{i-1})h\bigr]^2}{2\sigma^2x_{i-1}h}\biggr\}.
\]
We now wish to derive sufficient summary statistics for the parameters of interest, based on the complete loglikelihood.
Regarding $\sigma^2_{\varepsilon}$ this is trivial as we only have to consider $\sum_{j=1}^n \log p(y_j|x_j;\theta)$ to find that a sufficient statistic is $S_{\sigma^2_{\varepsilon}}=\sum_{j=1}^n(y_j-x_j)^2$. Regarding the remaining parameters we have to consider $\sum_{i=1}^N\log p(x_i|x_{i-1};\btheta)$. For $\sigma^2$ it is clear that a sufficient statistic is
\[
S_{\sigma^2} = \sum_{i=1}^N\biggl(\frac{\bigl[x_i-x_{i-1}-(\frac{Dose\cdot K_a \cdot K_e}{Cl}e^{-K_a \tau_{i-1}}-K_ex_{i-1})h\bigr]^2}{x_{i-1}h}\biggr).
\]
Regarding $K_e$ and $Cl$ things are a bit more complicated: we can write
\begin{align*}
\sum_{i=1}^N \log p(x_i|x_{i-1};\btheta)
&\propto  \sum_{i=1}^N \frac{\bigl[x_i-x_{i-1}-(\frac{Dose\cdot K_a \cdot K_e}{Cl}e^{-K_a \tau_{i-1}}-K_ex_{i-1})h\bigr]^2}{x_{i-1}}\\
&= \sum_{i=1}^N \biggl[\frac{x_i-x_{i-1}}{\sqrt{x_{i-1}}}-\biggl(\frac{Dose\cdot K_a \cdot K_e}{Cl{\sqrt{x_{i-1}}}}e^{-K_a \tau_{i-1}}-\frac{K_ex_{i-1}}{\sqrt{x_{i-1}}}\biggr)h\biggr]^2.
\end{align*}
The last equality suggests a linear regression approach $E(V)=\beta_1C_1+\beta_2C_2$ for ``responses'' $V_i=(x_i-x_{i-1})/\sqrt{x_{i-1}}$ and ``covariates'' 
\begin{align*}
C_{i1}&=\frac{Dose\cdot K_a e^{-K_a\tau_{i-1}}h}{\sqrt{x_{i-1}}}\\
C_{i2}&= -\frac{x_{i-1}}{\sqrt{x_{i-1}}}h=-\sqrt{x_{i-1}}h
\end{align*}
and $\beta_1=K_e/Cl$, $\beta_2=K_e$. By considering the design matrix $\bC$ with columns $\bC_1$ and $\bC_2$, that is $\bC = [\bC_1, \bC_2]$, from standard regression theory we have that $\hat{\boldsymbol{\beta}}=(\bC'\bC)^{-1}\bC'\bV$ is a sufficient statistic for $\boldsymbol{\beta}=(\beta_1,\beta_2)$, where $'$ denotes transposition. We take $S_{K_e}:=\hat{\beta}_2$ also to be used as the updated value of $K_e$ in the maximisations step of SAEM. Then we have that $\hat{\beta}_1$ is sufficient for the ratio $K_e/Cl$ and use $\hat{\beta}_2/\hat{\beta}_1$ as the update of $Cl$ in the M-step of SAEM. The updated values of $\sigma$ and $\sigma_{\varepsilon}$ are given by $\sqrt{S_{\sigma^2}/N}$ and $\sqrt{S_{\sigma^2_{\varepsilon}}/n}$ respectively.

\end{document}